\documentclass[12pt,preprint]{aastex}

\usepackage{lineno}

\slugcomment{}

\shorttitle{The Origin of the EGB}
\shortauthors{Ajello et al.}

\usepackage{calrsfs,euscript,mathrsfs,amssymb}
\usepackage{amsmath}
\begin{document}


\title{The Origin of the Extragalactic Gamma-Ray Background and
Implications for Dark-Matter Annihilation}



\author{
M.~Ajello\altaffilmark{1}, 
D.~Gasparrini\altaffilmark{2,3}, 
M.~S\'anchez-Conde\altaffilmark{4,5,6}, 
G.~Zaharijas\altaffilmark{7,8,9}
M.~Gustafsson\altaffilmark{10,11}, 
J.~Cohen-Tanugi\altaffilmark{12}, 
C.~D.~Dermer\altaffilmark{13}, 
Y.~Inoue\altaffilmark{14},
D.~Hartmann\altaffilmark{1}, 
M.~Ackermann\altaffilmark{15}, 
K.~Bechtol\altaffilmark{16}, 
A.~Franckowiak\altaffilmark{4},
A.~Reimer\altaffilmark{17}, 
R.~W.~Romani\altaffilmark{4}, 
A.~W.~Strong\altaffilmark{18}
}
\altaffiltext{1}{Department of Physics and Astronomy, Clemson University, Kinard Lab of Physics, Clemson, SC 29634-0978, USA}
\altaffiltext{2}{Agenzia Spaziale Italiana (ASI) Science Data Center, I-00133 Roma, Italy}
\altaffiltext{3}{INAF Osservatorio Astronomico di Roma, I-00040 Monte Porzio Catone (Roma), Italy}
\altaffiltext{4}{W. W. Hansen Experimental Physics Laboratory, Kavli Institute for Particle Astrophysics and Cosmology, Department of Physics and SLAC National Accelerator Laboratory, Stanford University, Stanford, CA 94305, USA}
\altaffiltext{5}{Department of Physics, Stockholm University, AlbaNova, SE-106 91 Stockholm, Sweden}
\altaffiltext{6}{The Oskar Klein Centre for Cosmoparticle Physics, AlbaNova, SE-106 91 Stockholm, Sweden}
\altaffiltext{7}{Istituto Nazionale di Fisica Nucleare, Sezione di Trieste, and Universit\`a di Trieste, I-34127 Trieste, Italy}
\altaffiltext{8}{The Abdus Salam International Center for Theoretical Physics, Strada Costiera 11, Trieste 34151 - Italy}
\altaffiltext{9}{Laboratory for Astroparticle Physics, University of Nova Gorica, Vipavska 13, SI-5000 Nova Gorica, Slovenia}
\altaffiltext{10}{Service de Physique Theorique, Universite Libre de Bruxelles (ULB),  Bld du Triomphe, CP225, 1050 Brussels, Belgium}
\altaffiltext{11}{Institut f\"ur Theoretische Physik, Friedrich-Hund-Platz 1, D-37077 G\"ottingen, Germany}

\altaffiltext{12}{Laboratoire Univers et Particules de Montpellier, Universit\'e Montpellier 2, CNRS/IN2P3, Montpellier, France}
\altaffiltext{13}{Space Science Division, Naval Research Laboratory, Washington, DC 20375-5352, USA}
\altaffiltext{14}{Institute of Space and Astronautical Science, Japan Aerospace Exploration Agency, 3-1-1 Yoshinodai, Chuo-ku, Sagamihara, Kanagawa 252-5210, Japan}
\altaffiltext{15}{Deutsches Elektronen Synchrotron DESY, D-15738 Zeuthen, Germany}
\altaffiltext{16}{Kavli Institute for Cosmological Physics, University of Chicago, Chicago, IL 60637, USA}

\altaffiltext{17}{Institut f\"ur Astro- und Teilchenphysik and Institut f\"ur Theoretische Physik, Leopold-Franzens-Universit\"at Innsbruck, A-6020 Innsbruck, Austria}
\altaffiltext{18}{Max-Planck Institut f\"ur extraterrestrische Physik, 85748 Garching, Germany}

\email{majello@clemson.edu, gasparrini@asdc.asi.it,
sanchezconde@fysik.su.se, gzaharijas@ung.si, mgustafs@ulb.ac.be}

%
%
%
%

\begin{abstract}
The origin of the extragalactic $\gamma$-ray background (EGB) 
 has been debated for some time. 
{ The EGB comprises the $\gamma$-ray emission from resolved and unresolved extragalactic sources, such as blazars, star-forming galaxies and radio galaxies, as well as radiation from truly diffuse processes.}
This letter focuses on the blazar source class, the most numerous detected population, and presents an updated luminosity function and spectral energy distribution model consistent with the blazar observations performed by the 
{\it Fermi} Large Area Telescope (LAT).
We show that blazars account for 50$^{+12}_{-11}$\,\% 
of the EGB photons ($>$0.1\,GeV),
 and that {\it Fermi}-LAT has already resolved $\sim$70\,\%
of this contribution. Blazars, and in particular low-luminosity
hard-spectrum nearby sources like BL Lacs, are responsible for most of the
EGB  emission above 100\,GeV. We find that the extragalactic background light, which attenuates blazars' high-energy emission, is responsible for the high-energy cut-off observed in the EGB spectrum.
Finally, we show that blazars, star-forming galaxies and radio galaxies can naturally account for the amplitude and spectral shape of the
background in the 0.1--820\,GeV range, leaving only modest room for other contributions. This allows us to set competitive constraints on  the dark-matter annihilation cross section.
\end{abstract}

\keywords{cosmology: observations -- dark matter --
 gamma rays: diffuse background -- galaxies: jets
galaxies: active --surveys}

%
%
\section{Introduction}

The Large Area Telescope \citep[LAT,][]{atwood09} on {\it Fermi}  has recently allowed a broadband, accurate, measurement
of the  extragalactic $\gamma$-ray background (EGB),
 the integrated emission of {\it all} resolved and unresolved
extragalactic GeV sources, characterizing its
intensity over almost 4 decades in energy between 0.1\,GeV and 
820\,GeV\citep[][ hereafter AC14]{lat_egb2}.
At these energies, the EGB spectrum is found  compatible with a power law 
with a photon index of 2.32($\pm0.02$) that is exponentially  cut off at 
279($\pm$52)\,GeV (AC14). 
Such cut-off, observed for the first time, may be caused by the
extragalactic background light \citep[EBL;][]{gould66,stecker92}.
Yet, this observation alone is not 
 sufficient to identify which process or 
source population is responsible for the EGB. Specifically,
the EGB may encompass the signatures of processes generating
a truly diffuse background, like intergalactic shocks 
\citep[e.g.][]{loeb00,miniati02}, $\gamma$-ray
emission induced by ultra-high energy cosmic rays in intergalactic space \citep[see e.g.][]{bhattacharjee00},  and
dark matter (DM) annihilation \citep[e.g.][]{Ullio02}.

Besides truly diffuse processes \citep{ahlers11}, unresolved point-like sources might be responsible
for a substantial part of the EGB \citep[][]{dermer07}. 
At high Galactic latitudes, {\it Fermi}-LAT has
detected blazars, star-forming galaxies, radio galaxies and millisecond 
pulsars \citep[][]{1fgl,2fgl}. Extensive analyses were recently performed 
to assess the contribution of all these source classes to the EGB.
Blazars, which constitute the largest 
population detected by the LAT, were found to contribute $\sim$20--30\,\%
of the unresolved EGB \citep{pop_pap,singal12,harding12,dimauro14}, 
and a larger fraction in some models \citep[][]{stecker11}.

Star-forming galaxies produce
 $\gamma$-rays in cosmic-ray interactions, with the acceleration of cosmic rays
 ultimately powered
by star formation \citep{thompson07,lacki14}. Being dimmer but more
numerous than blazars, star-forming galaxies
 might be responsible for 10--30\,\% 
of the 0.1--100\,GeV EGB photons 
 \citep{fields10,makiya2010,lat_starforming}. A similar
argument holds for misaligned AGN (e.g. radio galaxies), which
were recently found to produce $\sim$20\,\% of the EGB
\citep{inoue11b,dimauro13}. 

However, large uncertainties remain for  the contributions of the above source
classes. 
 In this paper we present (in $\S$~\ref{sec:analysis})
improved modeling  of the evolution
and of the spectral energy distributions (SEDs) of blazars, which allows
us to quantify their integrated emission.
We show (in $\S$~\ref{sec:discussion}) that the  integrated emission
of blazars, star-forming and radio galaxies naturally accounts
for the amplitude and spectrum of the new EGB measurement 
over the entire 0.1--820\,GeV
energy range. We then use this information to place  constraints 
on  the DM annihilation cross section. 
Throughout this paper, we adopt
{ H$_0$=67\,km s$^{-1}$ Mpc$^{-1}$}, $\Omega_M$=1-$\Omega_{\Lambda}$=0.30.

%
%
\section{Analysis and Modeling}
\label{sec:analysis}

The contribution of blazars to the EGB was already estimated in \cite{pop_pap} by
extrapolating the $\log N$-$\log S$  below the {\it Fermi}-LAT detection
threshold. Taking advantage of recent follow up observations \cite[e.g.,][]{shaw13,ajello14}, 
{ we derive new models  for the luminosity and redshift evolution 
of the whole blazar class and of its SED. In this section, 
these models are constrained using 
blazar data (fluxes, redshifts and photon indices) from \cite{agn_cat}, 
the $>$10\,GeV $\log N$-$\log S$
from \cite{1FHL} and information on the spectral curvature  
of blazars \citep{3LAC} to robustly estimate the integrated emission of blazars.
}

We rely on the sample of 403 blazars detected with test statistic\footnote{See
\cite{1fgl} for a definition.}  $>$50
 at $|b|>$15$^\circ$ in \cite{agn_cat}, for which a determination of the LAT detection efficiency exists \citep{pop_pap}.
This sample includes 211 BL Lacs, 186 flat-spectrum radio
quasars (FSRQs) and 6 blazars of uncertain type.
All but 109 BL Lacs have a spectroscopic redshift and
\cite{ajello14} provide redshift constraints for 104 (out of the 109)
BL Lacs.  
The fraction of sources with redshift information is $\sim$99\,\%
while the incompleteness of the sample, due to unassociated
sources that might be blazars, is $\sim$10\,\%. 
We do not separate the two blazar sub-classes
(FSRQs, and BL Lacs) 
because the larger sample allows  a better determination
of the integrated emission from the whole population,
{ and averages over the luminosity functions
of the two populations in the regime of overlapping luminosities. }
In order to derive the luminosity function (LF),
we use the  bootstrap Monte Carlo approach developed in \cite{ajello14}
that allows us to treat properly those sources with an imprecise redshift,
providing a robust error estimate.

We test models of primarily luminosity evolution (PLE),  
primarily density evolution (PDE)
and luminosity-dependent density evolution (LDDE). 
 In all these models 
the sources experience both luminosity and density evolution.
The LF at redshift $z=0$, for sources of { 0.1-100\,GeV rest-frame 
luminosity $L_{\gamma}$   (in erg s$^{-1}$)},
  is modeled as a double power law
multiplied by the photon index distribution:
\begin{equation}
\Phi(L_{\gamma},z=0, \Gamma) = \frac{{\rm d}N}{{\rm d}L_{\gamma}{\rm d}V{\rm d}\Gamma}= 
\frac{A}{\ln(10)L_{\gamma}} 
\left[\left(\frac{L_{\gamma}}{L_{*}}\right)^{\gamma_1}+
\left(\frac{L_{\gamma}}{L_{*}}\right)^{\gamma_2} 
\right]^{-1}\cdot  e^{-0.5[\Gamma-\mu(L_{\gamma})]^2/\sigma^2 } 
\label{eq:lf0}
\end{equation}
where $\mu$ and $\sigma$ are, respectively, the Gaussian mean and dispersion of
the photon  index ($\Gamma)$ distribution.
We allow  $\mu$ to change with luminosity as:
\begin{equation}
\label{eq:blazseq}
\mu(L_{\gamma}) = \mu^* + \beta \times [\log(L_{\gamma}) - 46].
\end{equation}
The evolution is parametrized by the `evolutionary factor' $e(z,L_{\gamma})$,
which is applied to the PDE and LDDE as:
\begin{equation}
\Phi(L_{\gamma},z,\Gamma) = \Phi(L_{\gamma},z=0,\Gamma) \times e(z,L_{\gamma}),
\label{eq:pde}
\end{equation}
and to the PLE as:
\begin{equation}
\Phi(L_{\gamma},z,\Gamma) = \Phi(L_{\gamma}/e(z,L_{\gamma}),z=0,\Gamma).
\label{eq:ple}
\end{equation}
For both the PLE and PDE the evolutionary factor is:
\begin{equation}
e(z,L_{\gamma}) = (1+z)^{k_d} e^{z/\xi},
\label{eq:ez}
\end{equation}
with 
\begin{equation}
k_d=k^* + \tau \times [\log(L_{\gamma})-46].
\label{eq:tau}
\end{equation}
For the LDDE\footnote{Note that in \cite{ajello14} the exponents 
$p_1$ and $p_2$ were reported with the wrong sign. See Eq.~\ref{eq:evol}
for the correct ones.} it is:

\begin{equation}
e(z,L_{\gamma})= \left[ 
\left( \frac{1+z}{1+z_c(L_{\gamma})}\right)^{-p_1(L_{\gamma})} + 
\left( \frac{1+z}{1+z_c(L_{\gamma})}\right)^{-p_2(L_{\gamma})} 	
 \right]^{-1},
\label{eq:evol}
\end{equation}

with 
\begin{equation}
z_c(L_{\gamma})= z_c^*\cdot (L_{\gamma}/10^{48})^{\alpha},
\label{eq:zpeak}
\end{equation}
\begin{equation}
p_1(L_{\gamma}) = p_1^* + \tau \times (\log(L_{\gamma})-46),
\label{eq:p1}
\end{equation}
\begin{equation}
p_2(L_{\gamma}) = p_2^* + \delta \times (\log(L_{\gamma})-46).
\label{eq:p2}
\end{equation}

\begin{figure*}[ht!]
  \begin{center}
  \begin{tabular}{cc}
\hspace{-1cm}
    \includegraphics[scale=0.45]{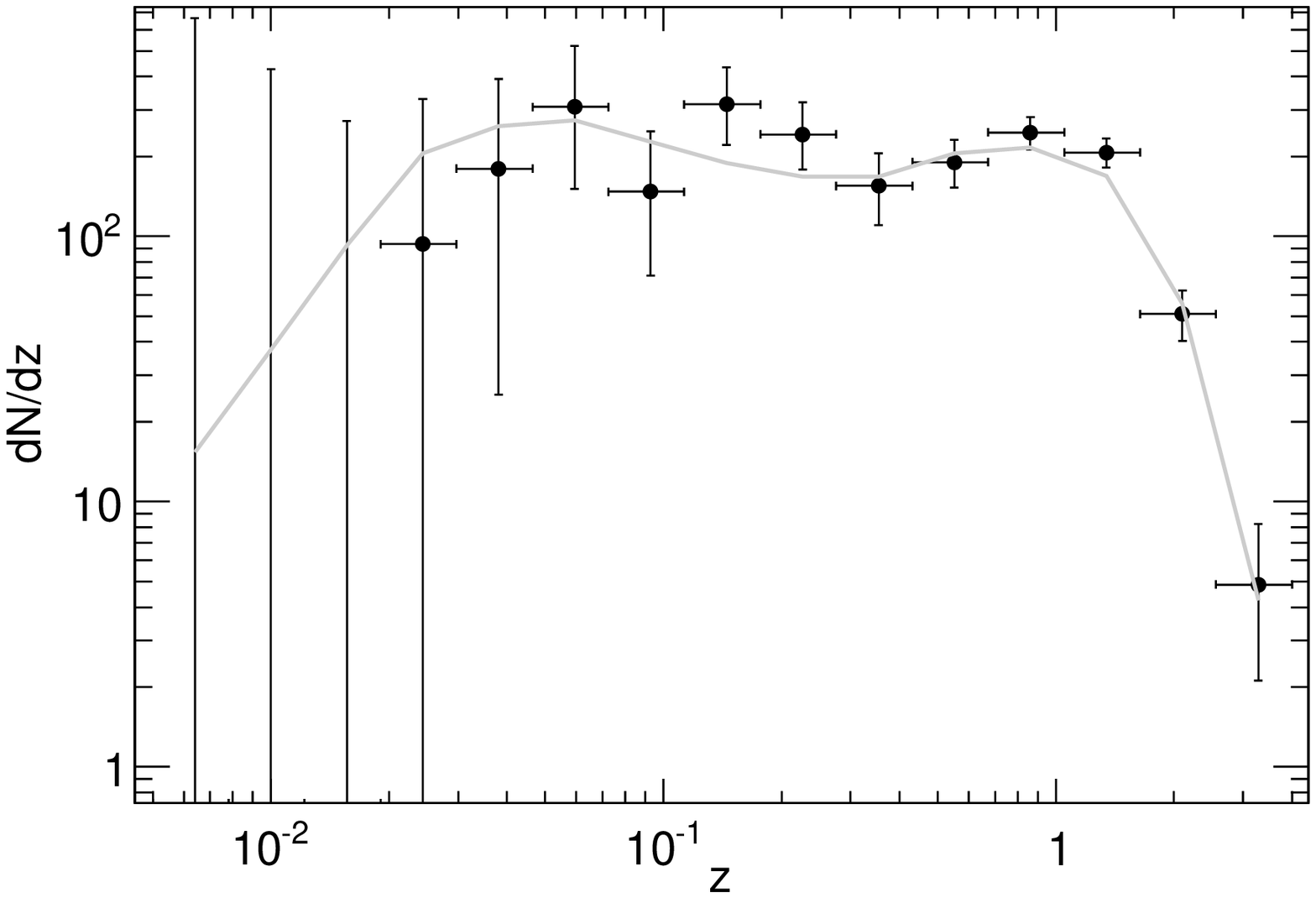} &
\hspace{-1cm}
  	 \includegraphics[scale=0.45]{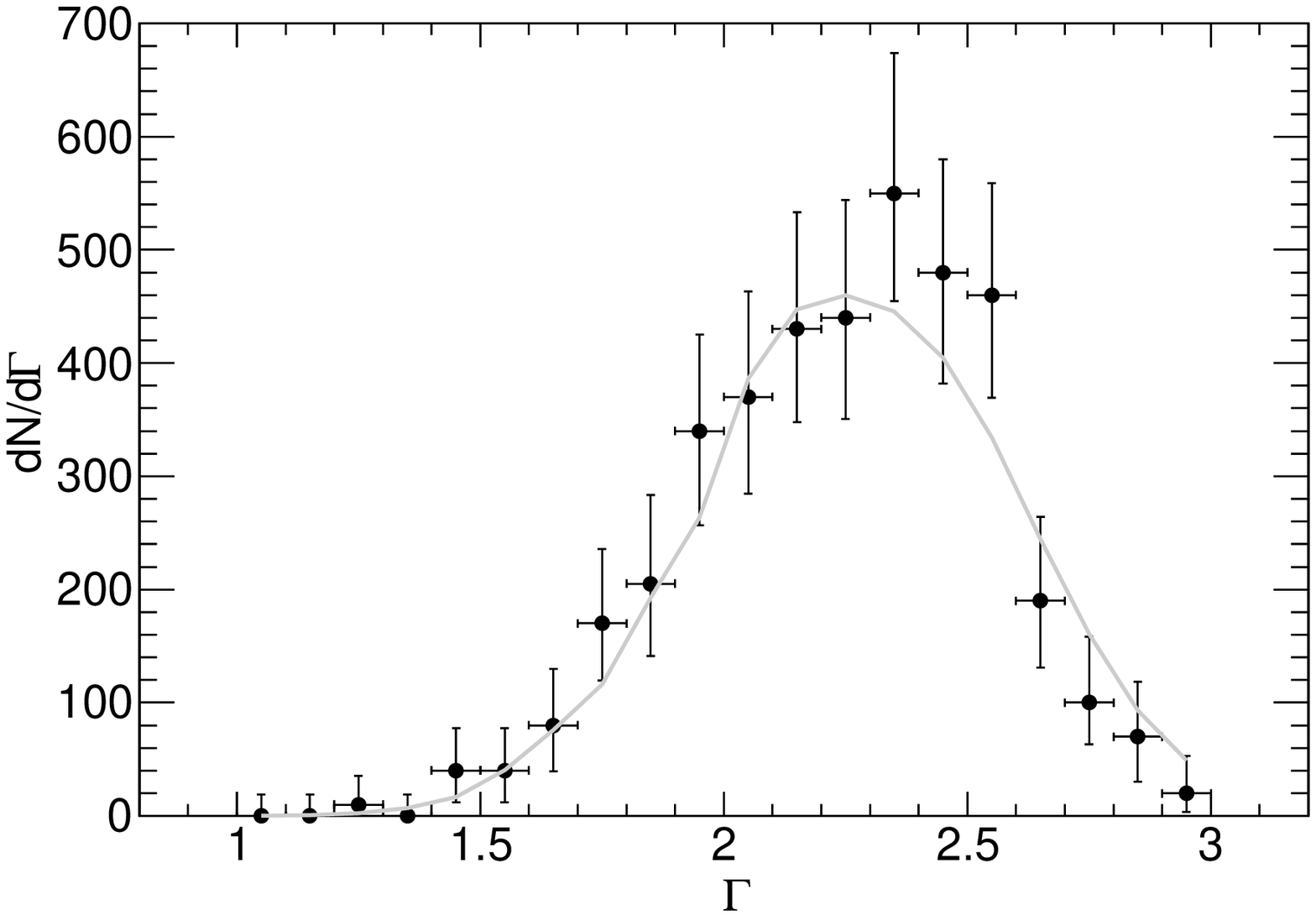} \\
\hspace{-1cm}
 	 \includegraphics[scale=0.45]{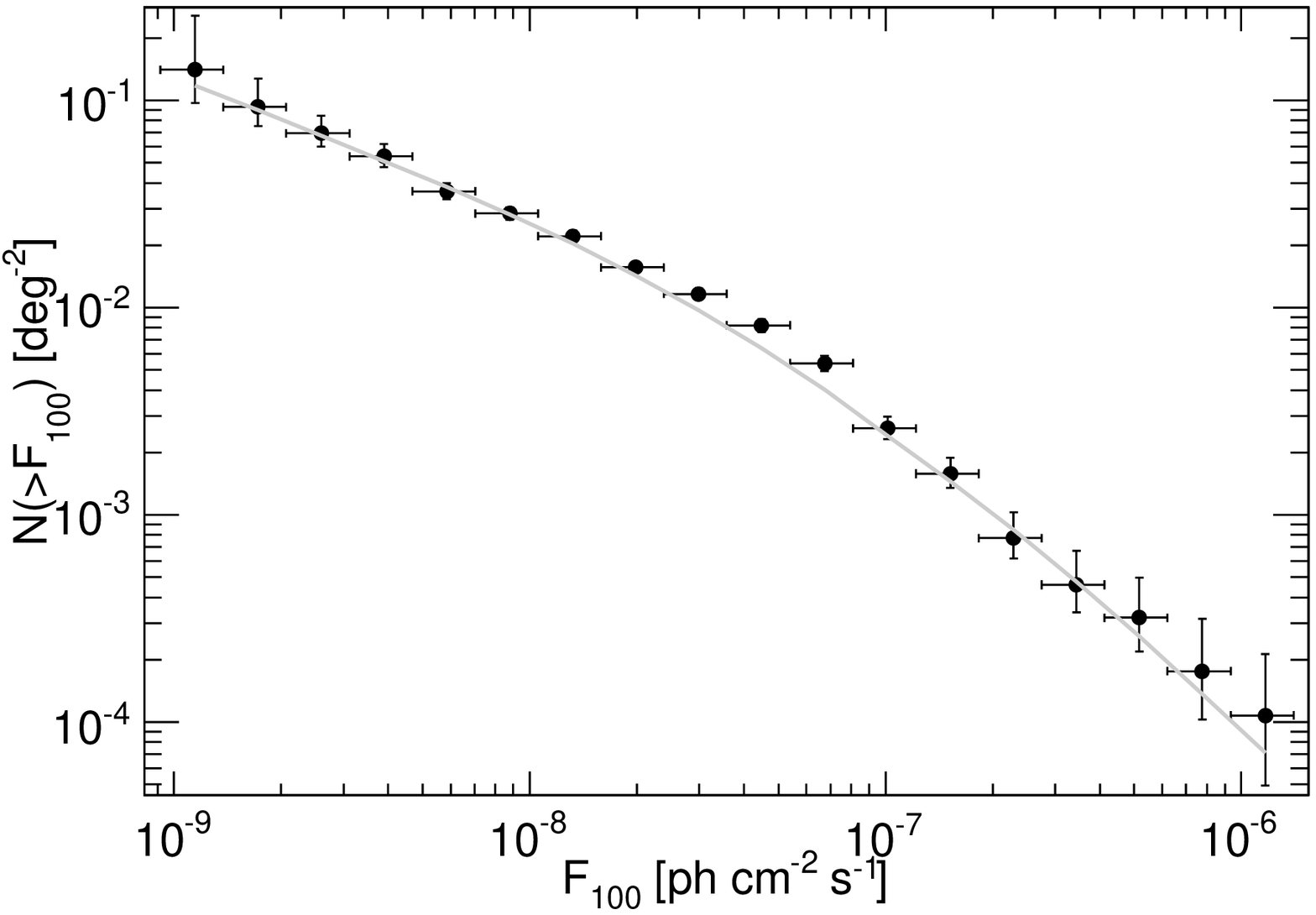} &
\hspace{-1cm}
	 \includegraphics[scale=0.45]{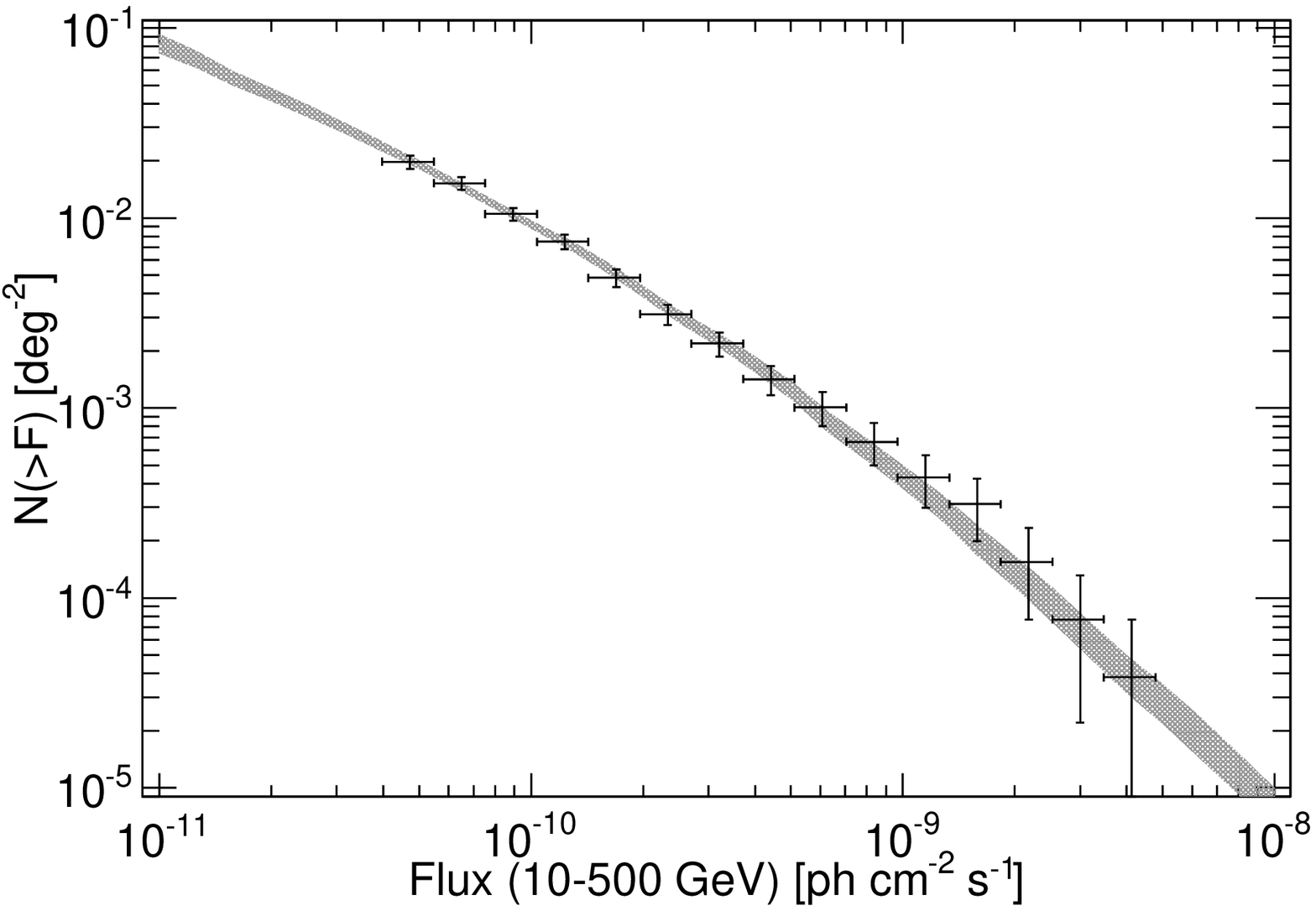}
\end{tabular}
  \end{center}
\caption{Observed redshift (upper left), photon index (upper right), 
0.1--100\,GeV source-count (lower left), and 10--500\,GeV source-count
 (lower right)  distributions of {\it Fermi}-LAT blazars.
For the upper panels, 
the continuous solid line is  the PLE model convolved
with the detection efficiency of {\it Fermi}-LAT \citep[see][]{pop_pap},
while for the lower ones
it represents the predictions of the LF models. The 68\,\% uncertainty
band in the lower right panel shows the prediction, for the 10--500\,GeV source counts,  of the LF and SED model.
Error bars compatible with zero are 1\,$\sigma$ upper limits for the case
of observing zero events in a given bin.
\label{fig:pde}}
\end{figure*}

All the model parameters reported in Eqs.\,1-10 are fitted,
through a maximum likelihood unbinned algorithm 
\cite[see \S~3 in][]{ajello14}, to the {\it Fermi}-LAT  
($L_{\gamma}$, $z$, $\Gamma$)  data (see Tab.~\ref{tab:lf}).
Among the three LF models, the LDDE model produces the largest log-likelihood,
  however a simple likelihood ratio test cannot 
be used to compare these non-nested models.
We find that all three LFs
provide an acceptable description of the
LAT data (flux, luminosity and photon indices, see Fig.~\ref{fig:pde}),
and more importantly predict
{ comparable} levels for the blazar  integrated emission 
(see Tab.~\ref{tab:lf}).

Blazars are known to have curved spectra when observed over a few decades 
in energy.
It is thus important to have
a reliable model of the high-energy component of the blazar SEDs.
Here we use a double power-law model attenuated by the EBL:
\begin{equation}
\frac{dN_{\gamma}}{dE}  =   K\left[ 
\left(\frac{E}{E_b} \right)^{\gamma_{a}} +
\left(\frac{E}{E_b} \right)^{\gamma_{b}} \right]^{-1}
\cdot e^{-\tau(E,z)} \ \ \ \ [{\rm ph\ cm^{-2} s^{-1} GeV^{-1}}]
\label{eq:sed}
\end{equation}
We  rely on the EBL model of \cite{finke10}, and  
use $\gamma_a$=1.7 and $\gamma_b$=2.6, which 
reproduces the
long-term averaged spectra 
of bright BL Lacs with GeV-TeV measurements 
\citep[RBS 0413, Mrk 421, Mrk 501, see][]{rbs0413,lat_421,lat_501} and those of bright FSRQs (like 
3C 454.3, 3C 279, 3C 273, etc.)  observed by {\it Fermi}-LAT. 
Typically, all blazar spectra show { a high-energy cut-off}
 that reflects the distributions  of the accelerated particles.
This is located at $E\geq$1\,TeV and $E\leq$100\,GeV for BL Lacs
and FSRQs respectively.
Here, 
including such cut-offs makes very little difference because for BL Lacs
the cut-offs are at energies larger than those { probed here}, while
{ for FSRQs, because of the larger average redshifts,  the EBL efficiently suppresses their $>$50\,GeV flux.}

\begin{figure*}[ht!]
  \begin{center}
  \begin{tabular}{cc}
\hspace{-1cm}
    \includegraphics[scale=0.46]{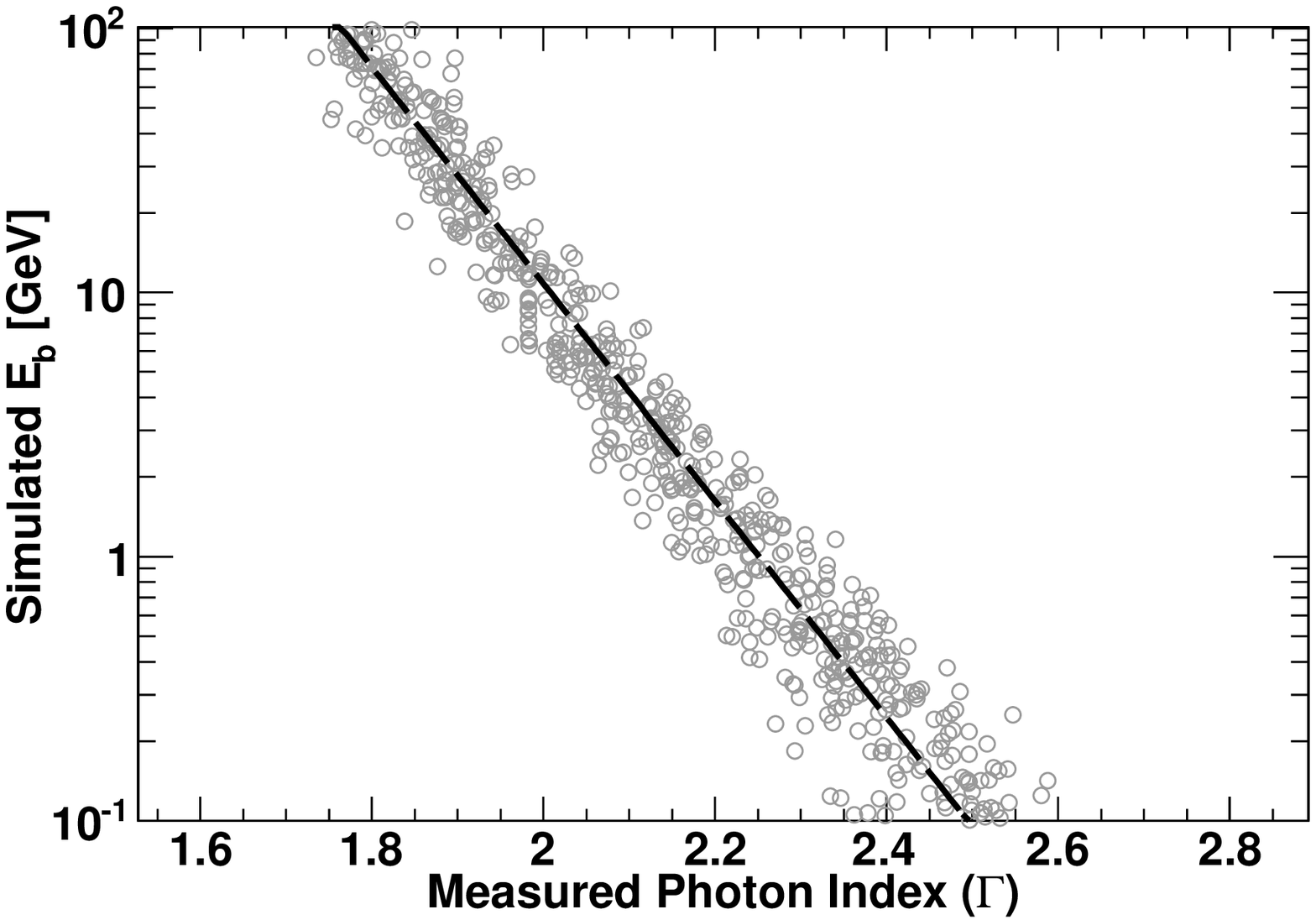} &
\hspace{-1cm}
  	 \includegraphics[scale=0.46]{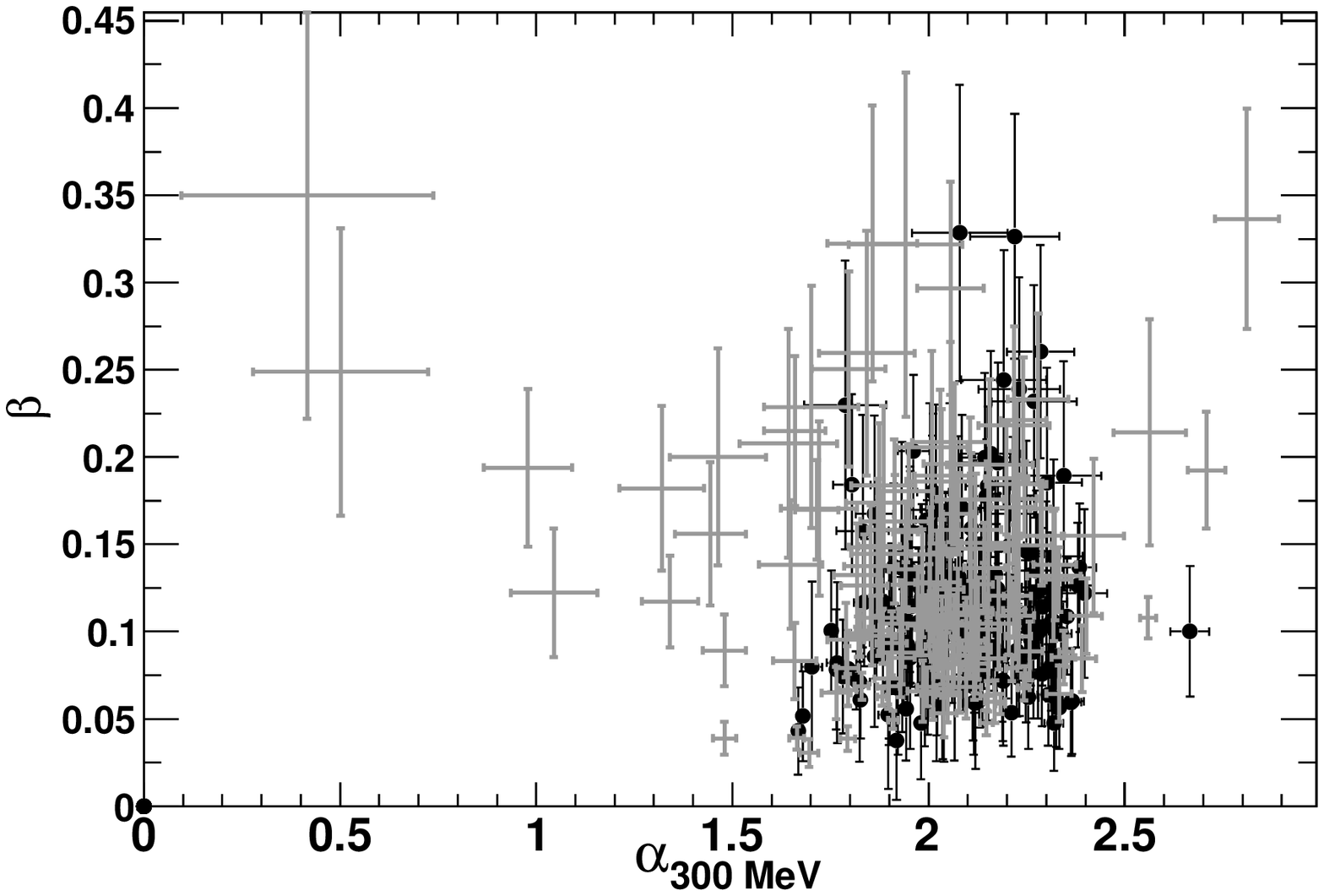} \\
\end{tabular}
  \end{center}
\caption{Left Panel: Simulated break energy $E_b$ (for Eq.~\ref{eq:sed}
with $\gamma_a$=1.7, $\gamma_b$=2.6) versus measured power-law 
photon index for a set of simulated  blazars.
The dashed line represents the best fit described in the text.
Right Panel: photon index ($\alpha$, at 300\,MeV) 
and curvature $\beta$ (black data points) of the best-fitting
logParabola models to simulated double power-law spectra
(eg. Eq.~\ref{eq:sed} with $\gamma_a$=1.7 and $\gamma_b$=2.6).
The gray datapoints show the 
parameters for all the blazars (184)  whose curvature is
significantly detected in the 3LAC catalog \citep{3LAC}.
\label{fig:sed}}
\end{figure*}

For the model reported above, the  high-energy peak
 is a function of $E_b$ alone, for fixed $\gamma_a$ and  $\gamma_b$.
We calibrated the relationships between  $E_{b}$ and the
LAT-measured  power-law photon index via simulations and found that it
can be approximated as $\log E_{b}(\rm GeV)\approx9.25 - 4.11\Gamma$
(see left panel of Fig.~\ref{fig:sed}). { The spectral curvature
seen in bright LAT blazars is typically characterized using a logParabola model  $dN/dE \propto (E/E_0)^{-\alpha - \beta {\rm log}(E/E_0)}$
(known to approximate blazar SED well only around their peak), where $\alpha$
is the photon index at energy $E_0$ and $\beta$ is the curvature parameter \citep{2fgl}. In order to ascertain that our SED model reproduces the correct amount
of spectral curvature observed in blazars, we simulated LAT observations
of $\sim$1600 blazars with fluxes randomly extracted from the 3LAC catalog
and a spectrum described by Eq.~\ref{eq:sed}. We treated these spectra
as the real data 
 and whenever the logParabola model was preferred
over the power law at $\geq$4\,$\sigma$ \citep[as in][]{2fgl} we estimated the 
$\alpha$ and $\beta$ parameters. As Fig.~\ref{fig:sed} (right panel) shows
these are found to be in good agreement with the parameters of the
real blazar set, validating our choice of the SED model.}

We thus use the above $E_{b}-\Gamma$  relation  
{ to predict}
the integrated emission of the blazar class that we compute as:
\begin{equation}
F_{EGB} (E_{\gamma}) = \int\limits^{\Gamma_{\rm max}=3.5}\limits_{\Gamma_{\rm min}=1.0}{\rm d}\Gamma \int\limits^{z_{\rm max}=6}\limits_{z_{\rm min}=10^{-3}}{\rm d}z
\int\limits^{L_{\gamma}^{\rm max}=10^{52}}\limits_{L_{\gamma}^{\rm min}=10^{43}}{\rm d}L_{\gamma} \cdot \Phi(L_{\gamma},z,\Gamma)
\cdot \frac{dN_{\gamma}}{dE}\cdot \frac{dV}{dz} 
\ \ [{\rm ph\ cm^{-2} s^{-1} sr^{-1} GeV^{-1}}],
\label{eq:egb}
\end{equation}
where $\Phi(L_{\gamma},z,\Gamma)$
and $\frac{dN_{\gamma}}{dE}$ are the LF and the spectrum reported above. 
Because the LF displays steep power laws at high
redshift and luminosity, the only limit that matters is $L_{\gamma}^{min}$,
which we set as the lowest observed luminosity. 
The normalization factor $K$ of Eq.~\ref{eq:sed} is chosen so
that a source  at redshift $z$ and with index $\Gamma$,
implying $E_b=E_b(\Gamma)$ given by the slope of Fig.~\ref{fig:sed} (left panel), has a rest-frame luminosity $L_{\gamma}$.
We also make sure that both the LF and SED models
are able to reproduce the 10--500\,GeV source counts \citep{1FHL}, which
is important to obtain a robust estimate of the contribution of blazars
to the high-energy EGB (see Fig.~\ref{fig:pde}). 

Integrating Eq.~\ref{eq:egb} above 0.1\,GeV
{ for the three LF models and averaging\footnote{We used a
weighted average with 1/$\sigma_i^2$ (e.g. inverse of 
flux variance for each model) weights.}} the results
yields that all blazars (including 
the resolved ones) emit 5.70($\pm1.06)\times10^{-6}$\,ph cm$^{-2}$ s$^{-1}$
sr$^{-1}$, where the error is dominated by the systematic uncertainties
(all similar in magnitude)
on the {\it Fermi}-LAT detection efficiency \citep[][]{pop_pap},
on the missing associations, the differences between the three
LF models, and the scatter of the $E_b-\Gamma$ relation.
When comparing this to the total EGB intensity  of 
11.3$^{+1.6}_{-1.5}\times10^{-6}$\,ph cm$^{-2}$ s$^{-1}$ sr$^{-1}$ (AC14) 
 we conclude that
blazars produce 50$^{+12}_{-11}$\,\% 
of the total EGB.
Since the resolved component of the EGB is 4.1($\pm0.4$)$\times10^{-6}$\,ph cm$^{-2}$ s$^{-1}$ sr$^{-1}$  (see AC14), and most of the detected
sources are blazars,
we conclude that {\it Fermi}-LAT has already resolved $\sim$70\,\% of the
total blazar emission.

\begin{figure*}[ht!]
  \begin{center}
  \begin{tabular}{c}
    \includegraphics[scale=0.7]{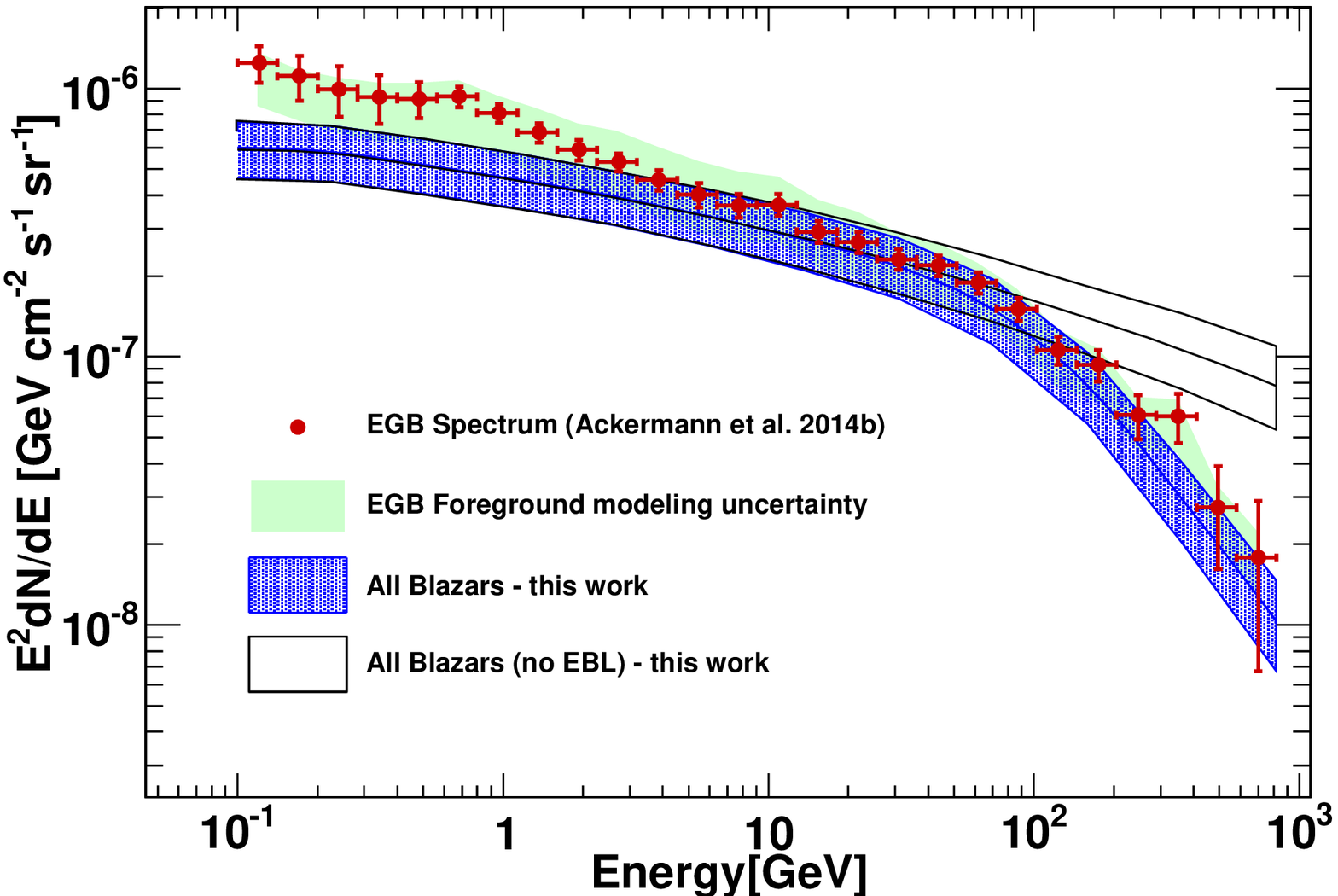} \\
  	 \includegraphics[scale=0.7,clip=true,trim=0 0 0 20]{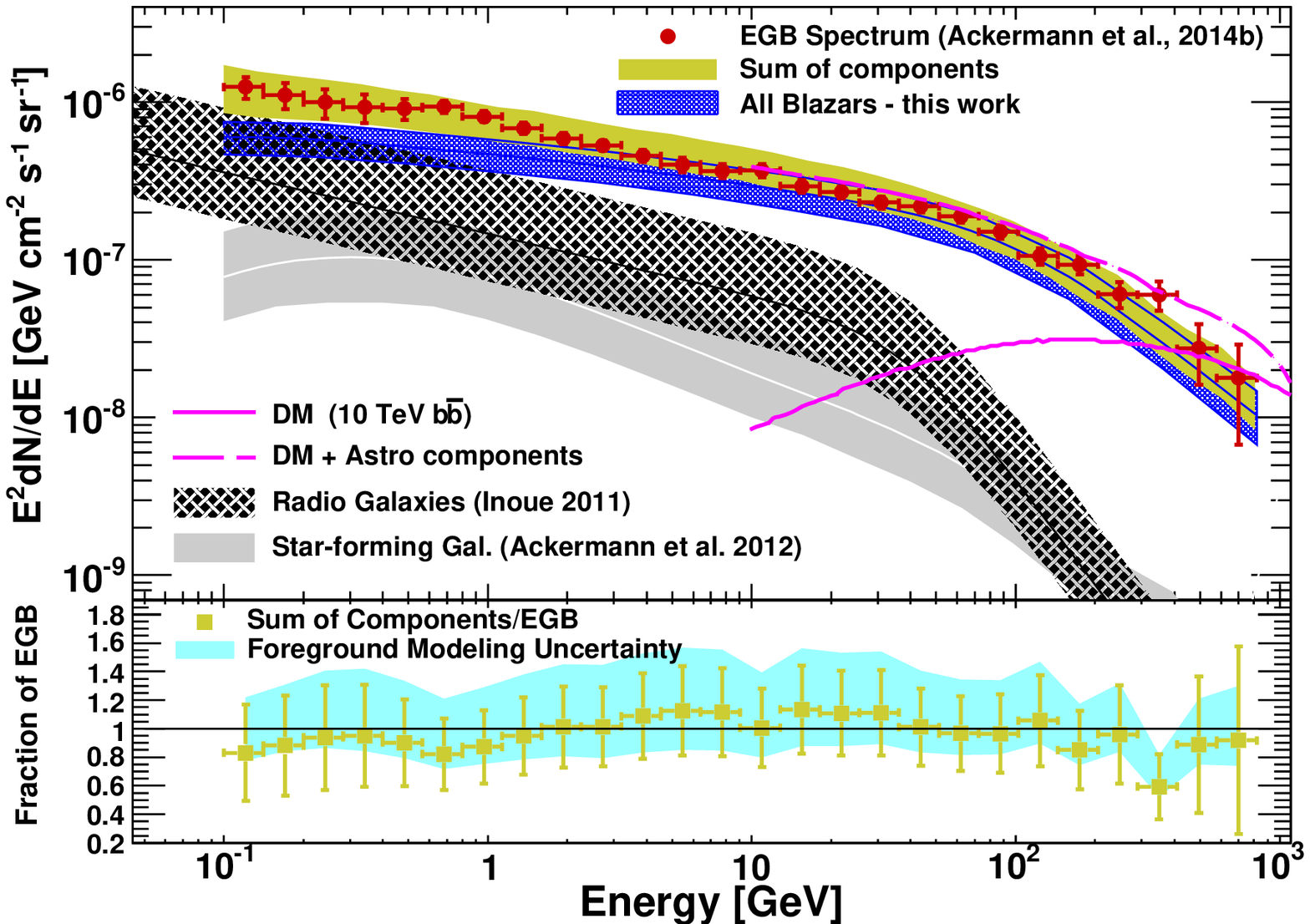} 
\end{tabular}
  \end{center}
\vspace{-1cm}
\caption{Top Panel: Integrated emission of blazars (with
and without EBL absorption), compared to the intensity of the EGB   
(datapoints from AC14).
Lower Panel: as above, but including
also the emission from star-forming galaxies \citep[gray band,][]{lat_starforming} and radio galaxies  \citep[black striped band,][]{inoue11b}  as
well as the sum of all non-exotic 
components (yellow band). { An example of 
DM-induced $\gamma$-ray signal ruled out by our analysis is shown
by the solid pink line, and summed with the non-exotic components (long-dashed pink line).}
 The inset shows the residual emission, computed as the ratio of the summed contribution to the EGB spectrum, as a function of energy as well as the uncertainty due to the foreground
emission models (see AC14).
\label{fig:egb}}
\end{figure*}

\begin{deluxetable}{lcccccccccccccc}
\tablewidth{-2pt}
\tabletypesize{\tiny}
\rotate
\tablecaption{Best-fit parameters of the LF models.
Parameter values were computed as the median of all the best-fit parameters
to the Monte Carlo sample, while the uncertainties (statistical only)
represent the 68\,\% containment regions around the median values.
{ Parameter names are reported  as ($k*$ or $p_1^*$)  and ($\xi$ or p$_2^*$)
for the PLE/PDE and LDDE models, respectively.}
\label{tab:lf}}
\tablehead{\colhead{Model}   & \colhead{Diffuse\tablenotemark{a}} &
\colhead{A\tablenotemark{b}} & \colhead{$\gamma_1$} & 
\colhead{L$_*$\tablenotemark{c}}              & \colhead{$\gamma_2$} &
\colhead{$k^*$ or p$_1^*$}               & \colhead{$\tau$}     &
\colhead{$\xi$ or p$_2^*$ }          & \colhead{$\delta$} &    
\colhead{z$_c^*$}  & 
\colhead{$\alpha$\tablenotemark{d}} &
\colhead{$\mu^*$}              & \colhead{$\beta$}    & 
\colhead{$\sigma$}         
}
\startdata 

PDE & 5.86$^{+0.19}_{-0.49}$ & $1.22^{+1.68}_{-1.11}$ & $2.80^{+1.15}_{-0.25}$ & $0.44^{+4.15}_{-0.15}$ & $1.26^{+0.09}_{-0.08}$ & $12.14^{+2.10}_{-1.74}$ & $2.79^{+0.56}_{-1.30}$ & $-0.15^{+0.02}_{-0.03}$ & \nodata & \nodata & \nodata  & $2.22^{+0.02}_{-0.02}$ & $0.10^{+0.02}_{-0.02}$ & $0.28^{+0.02}_{-0.01}$ \\

PLE & 5.76$^{+0.50}_{-0.40}$  & $19.3^{+9.7}_{-7.2}$ & $3.19^{+0.51}_{-0.40}$ & $8.75^{+4.09}_{-2.42}$ & $1.14^{+0.07}_{-0.08}$ & $4.41^{+0.61}_{-0.64}$ & $0.91^{+0.13}_{-0.15}$ & $-0.43^{+0.05}_{-0.07}$& \nodata  & \nodata & \nodata & $2.22^{+0.03}_{-0.03}$ & $0.10^{+0.02}_{-0.02}$ & $0.28^{+0.02}_{-0.01}$ \\

LDDE & 5.41$^{+0.57}_{-0.44}$  & $196^{+255}_{-130}$ & $0.50^{+0.14}_{-0.12}$ & $1.05^{+2.18}_{-0.56}$ & $1.83^{+0.63}_{-0.35}$ & $3.39^{+0.89}_{-0.70}$ & $3.16^{+1.45}_{-0.76}$ & $-4.96^{+2.25}_{-4.76}$ & $0.64^{+1.65}_{-1.05}$ & $1.25^{+0.19}_{-0.17}$ & $7.23^{+2.17}_{-2.99}$ & $2.22^{+0.03}_{-0.02}$ & $0.10^{+0.02}_{-0.02}$ & $0.28^{+0.02}_{-0.01}$ \\

\enddata
\tablenotetext{a}{Integrated blazar emission (0.1--820\,GeV), in units of
{  $10^{-6}$\,ph cm$^{-2}$ s$^{-1}$ sr$^{-1}$}, obtained by integrating the LF model between
the limits reported in $\S$~\ref{sec:analysis}.}
\tablenotetext{b}{{ In units of $10^{-2}$\,Gpc$^{-3}$.}}
\tablenotetext{c}{In units of $10^{48}$\,erg s$^{-1}$ for the PDE and LDDE models
 while  units of $10^{46}$\,erg s$^{-1}$ for the PLE model.}
\tablenotetext{d}{{ In units of $10^{-2}$.}}
\end{deluxetable}

%
%
\section{Discussion}
\label{sec:discussion}

Fig.~\ref{fig:egb} shows the spectrum of the integrated emission
of  blazars\footnote{We neglected the secondary emission
due to electromagnetic cascades created by electron-positron pairs generated
in the interaction of $\gamma$-rays with the EBL.}.
We find that
 the cut-off detected in the EGB spectrum is well explained by EBL absorption
of the high-energy blazar emission.
Above 100\,GeV, the majority of the EGB  is produced by
nearby ($z$$\lesssim$0.5) low-luminosity hard-spectrum blazars.
 Below this energy, blazars cannot account for the entire EGB, in
agreement with  previous
 findings \citep[see][]{pop_pap} that, in the 0.1--100\,GeV band, {\it unresolved} blazars can account for only $\sim$20\,\% of the  
{\it unresolved} EGB  intensity.
Furthermore, it is difficult to accommodate a blazar population that produces 
a larger fraction of the $<$100\,GeV EGB, because of the constraint
placed by the level of the small-scale anisotropies of the $\gamma$-ray sky 
as measured by {\it Fermi} \citep{aniso2012}. A blazar population
that reproduces the 0.1--100\,GeV source-count data \citep[][]{pop_pap} can account for $\sim$100\,\% of the angular power \citep[][]{cuoco12},
but only for $\sim$20--30\,\% of the unresolved EGB.

Therefore, the remaining $\sim$50\,\% ($\sim$5.6$\times10^{-6}$\,ph cm$^{-2}$ s$^{-1}$ sr$^{-1}$) of the total EGB intensity (particularly at $<$100\,GeV) 
must be produced by other populations or emission mechanisms.
Star-forming galaxies and radio galaxies that, in addition
to blazars and millisecond pulsars\footnote{Millisecond pulsars were shown to produce 
a negligible ($<$1\,\%) fraction of the EGB \citep[e.g.][]{calore14}.}, are detected by {\it Fermi}-LAT,
meet this requirement. 
Both star-forming and radio galaxies 
were shown to contribute
10--30\,\% of the EGB emission \citep{fields10,makiya2010,lat_starforming,inoue11b,dimauro13}.
By summing the contribution of star-forming galaxies \citep{lat_starforming} and radio galaxies \citep{inoue11b} to the contribution of blazars
derived here (see Fig.~\ref{fig:egb}) we find that these three 
populations can naturally account for the intensity of the EGB
across  the 0.1--820\,GeV range sampled by {\it Fermi}-LAT.
 This scenario does not change
if we adopt different models for the emission of 
 star-forming and radio galaxies \citep[e.g.][]{makiya2010,fields10,dimauro13}.

This study shows that the source populations already detected by {\it Fermi}-LAT
can account for the entire measured EGB, leaving little room for other contributions. This  can be used
to constrain the emission from `yet undetected' populations or diffuse
processes. One of the most intriguing mechanisms that can produce a diffuse
$\gamma$-ray flux
 is the self-annihilation of DM present in the Universe. 
Indeed, if DM is composed of self-annihilating Weakly Interactive Massive Particles (WIMPs) with masses of a few dozens to hundreds of GeV \citep[see, e.g.,][for a review]{bertone05}, then a diffuse 
GeV background may be expected from annihilations in DM halos across all cosmic epochs. This cosmological DM annihilation would thus contribute to the measured EGB, potentially yielding valuable information about the dark sector. No hints of a DM detection have been claimed up to now using the EGB. Yet, competitive limits on the DM annihilation cross section have been derived in several studies { relying on } the EGB intensity \citep[e.g.][]{cosmowimp1,calore2013,cholis14} or the anisotropy level \citep{german14}.

\begin{figure*}[ht!]
  \begin{center}
  \begin{tabular}{c}
\includegraphics[scale=0.6]{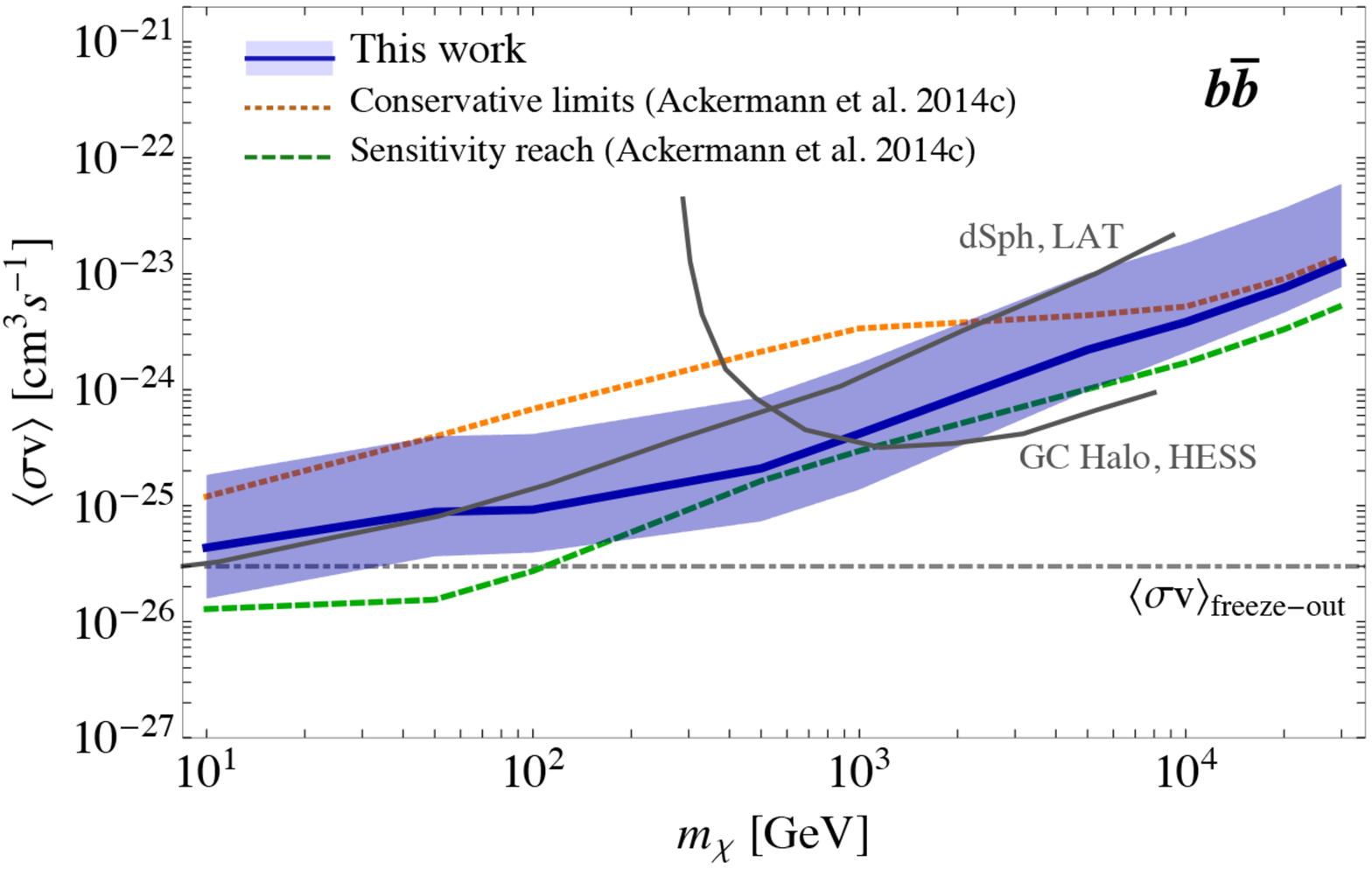} \\
\includegraphics[scale=0.6]{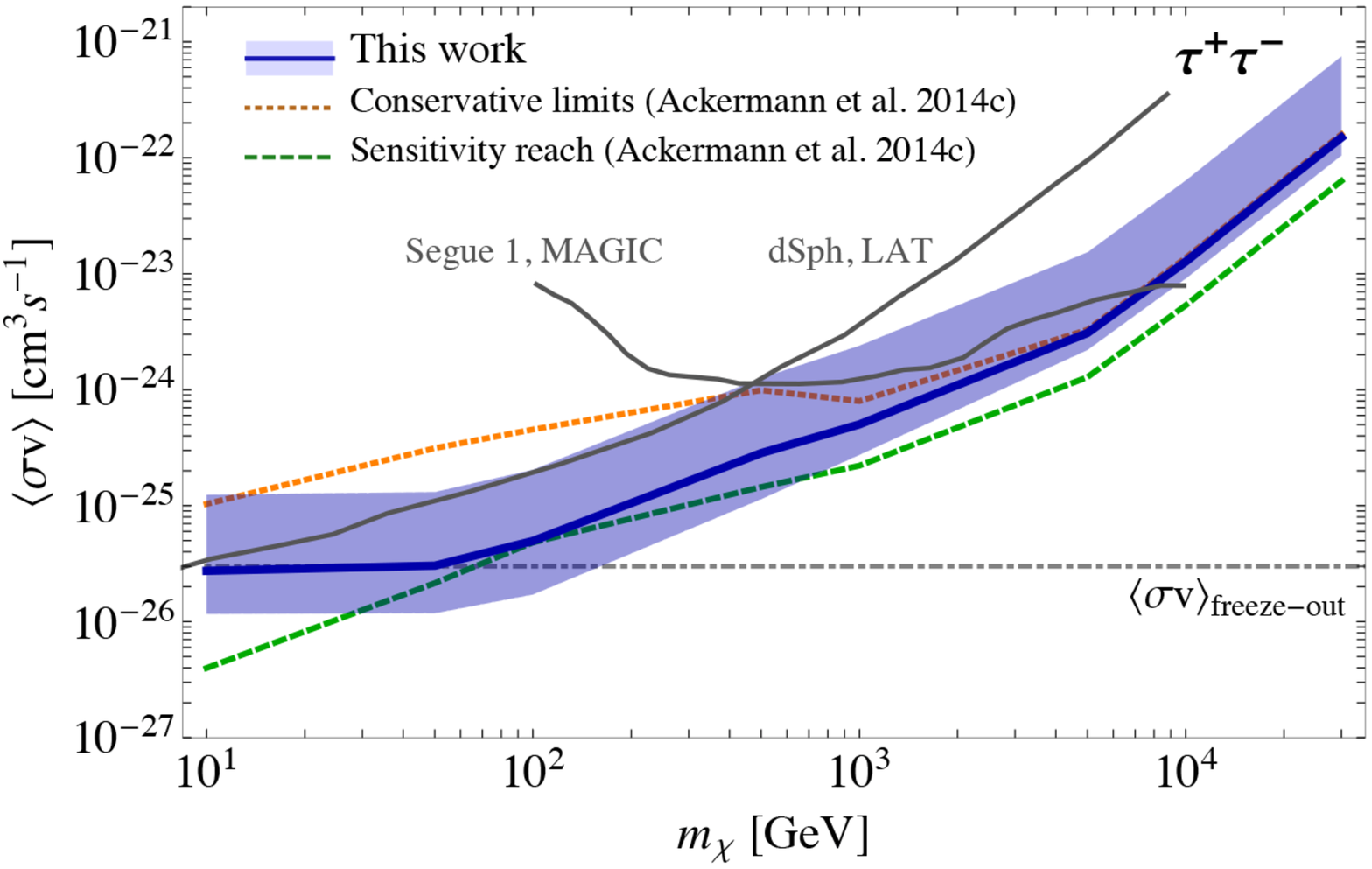} 
\end{tabular}
  \end{center}
\caption{
Upper limits on the self-annihilation cross section for the $b{\bar b}$ (top) and $\tau ^+ \tau ^-$ (bottom) channels as derived in this work (see $\S$~\ref{sec:discussion}) { compared to the conservative and sensitivity-reach  limits reported in  \cite{cosmowimp2}.
 The blue band reflects the range of the theoretical
predicted DM signal intensities, due to the uncertainties in the description
of DM subhalos in our Galaxy as well as other extragalactic
halos, adopting a cut-off minimal halo mass of 10$^{-6}$M$_\odot$.}
For comparison, limits reported in the literature are also shown \citep{hess_gc11,lat_dwarfs14,magic_segue14}.
\label{fig:DMlimits}}
\end{figure*}

Here, we use the main result of this analysis, 
that most of the EGB emission is produced by known source classes, 
to constrain the DM annihilation cross section.
{ We rule out DM models that, together with
point-like sources, overproduce the EGB emission 
at $\geq$2\,$\sigma$ level. This is achieved by defining:
\begin{equation}
\chi^2 = \underset{\mathcal{A}}{\mathrm{Min}}  \left[ \sum_i^N \frac{(F_{i,EGB} - \mathcal{\mathcal{A}}F_{i,ASTRO} - F_{i,DM})^2}{\sigma^2_i}+ \frac{(1-\mathcal{A})^2}{\sigma_{\mathcal{A}}^2}\right],
\label{eq:chi}
\end{equation}
where the sum runs over the $N$ bins of the EGB spectrum. 
$F_{i,EGB}, F_{i,ASTRO}, F_{i,DM}$ are the intensities of the EGB, point-like sources
and DM, $\mathcal{A}$ is a renormalization constant of the nominal 
integrated source intensity
and $\sigma_{\mathcal{A}}=\langle\sigma_{i,ASTRO}/F_{i,ASTRO}\rangle$
its average uncertainty. 
In Eq.~\ref{eq:chi}, {$\sigma_i$ is the sum (in quadrature)
 of the uncertainty on the unresolved EGB and the systematic uncertainty on the Galactic foreground (AC14).} 
We use the uncertainties on the unresolved EGB because the uncertainties on
the resolved source intensity are already taken into account in $\sigma_{\mathcal{A}}$.
The 2\,$\sigma$ limits are found when the DM signal worsens the $\chi^2$ by $\geq$4 with respect to the optimized $\chi^2$ with a free DM signal normalization (and a free $\mathcal{A}$). 
Following \cite{cosmowimp2}, predictions of the cosmological annihilation signal were obtained using both the {\it halo model} \citep{Ullio02,fornasa13} and the {\it power spectrum} approach \citep{serpico2011,sefusatti14}. 
{Though Eq.~\ref{eq:chi} neglects bin-to-bin correlations, we verified that our DM limits are within 10\,\% of those obtained if, for each DM signal,
we adopt the foreground model (from AC14) that gives the most conservative upper limit.}

An example of a ruled-out DM signal is reported in Fig.~\ref{fig:egb},
while Fig.~\ref{fig:DMlimits} shows the limits for DM annihilating to $b{\bar b}$ and $\tau^+ \tau^-$ channels including their uncertainties
due to the level of sub-halos in our Galaxy  and in all DM halos
\citep{masc14,cosmowimp2}. Our limits are compared
to the {\it conservative}  and {\it sensitivity-reach} limits reported in \cite{cosmowimp2}. The former assumes that the unresolved EGB is entirely due to DM annihilations, while the latter assumes the EGB is entirely produced by point-like sources and the DM annihilation is limited to what the uncertainties on the unresolved EGB allow. These represent extreme cases,   while our limits  represent a more realistic scenario\footnote{The {\it sensitivity-reach} limits in \cite{cosmowimp2} neglect the uncertainties in the integrated emission 
{ of non-exotic}
sources which, once taken into account, will weaken the constraints on the cross section.}, which, as expected, falls in between. Overall, our limits are 2-3 times more constraining than the conservative ones in \cite{cosmowimp2} thanks to the refined knowledge of the integrated emission from point-like sources derived here. 

This work shows that an analysis of the EGB and its components can  constrain diffuse emission mechanisms like  DM annihilation.  The comparison
of our limits with the {\it sensitivity-reach} limits
 of \cite{cosmowimp2} shows that reducing the overall uncertainties
is key to placing tighter constraints  on DM annihilation. This can be achieved by refining the estimate of the emission from star-forming and radio galaxies.}

\acknowledgments
The \textit{Fermi}-LAT Collaboration acknowledges support for LAT development, operation and data analysis from NASA and DOE (United States), CEA/Irfu and IN2P3/CNRS (France), ASI and INFN (Italy), MEXT, KEK, and JAXA (Japan), and the K.A.~Wallenberg Foundation, the Swedish Research Council and the National Space Board (Sweden). Science analysis support in the operations phase from INAF (Italy) and CNES (France) is also gratefully acknowledged.

{\it Facilities:} \facility{Fermi/LAT}

\bibliographystyle{apj}

\begin{thebibliography}{51}
\expandafter\ifx\csname natexlab\endcsname\relax\def\natexlab#1{#1}\fi

\bibitem[{{Abdo} {et~al.}(2010{\natexlab{a}}){Abdo}, {Ackermann}, {Ajello},
  {Baldini}, {Ballet}, {Barbiellini}, {Bastieri}, {Bechtol}, {Bellazzini},
  {Berenji}, {Blandford}, {Bloom}, {Bonamente}, {Borgland}, {Bouvier},
  {Bregeon}, {Brez}, {Brigida}, {Bruel}, {Burnett}, {Buson}, {Caliandro},
  {Cameron}, {Caraveo}, {Carrigan}, {Casandjian}, {Cecchi}, {{\c C}elik},
  {Chekhtman}, {Cheung}, {Chiang}, {Ciprini}, {Claus}, {Cohen-Tanugi},
  {Conrad}, {Cutini}, {Dermer}, {de Angelis}, {de Palma}, {Digel}, {Silva},
  {Drell}, {Dubois}, {Dumora}, {Edmonds}, {Farnier}, {Favuzzi}, {Fegan},
  {Focke}, {Fortin}, {Frailis}, {Fukazawa}, {Fusco}, {Gargano}, {Gasparrini},
  {Gehrels}, {Germani}, {Giglietto}, {Giordano}, {Glanzman}, {Godfrey},
  {Grove}, {Guillemot}, {Guiriec}, {Gustafsson}, {Hadasch}, {Harding}, {Horan},
  {Hughes}, {Johnson}, {Johnson}, {Kamae}, {Katagiri}, {Kataoka}, {Kawai},
  {Kerr}, {Kn{\"o}dlseder}, {Kuss}, {Lande}, {Latronico}, {Llena Garde},
  {Longo}, {Loparco}, {Lott}, {Lovellette}, {Lubrano}, {Makeev}, {Mazziotta},
  {McEnery}, {Meurer}, {Michelson}, {Mitthumsiri}, {Mizuno}, {Monte},
  {Monzani}, {Morselli}, {Moskalenko}, {Murgia}, {Nolan}, {Norris}, {Nuss},
  {Ohsugi}, {Omodei}, {Orlando}, {Ormes}, {Paneque}, {Panetta}, {Parent},
  {Pelassa}, {Pepe}, {Pesce-Rollins}, {Piron}, {Rain{\`o}}, {Rando}, {Reimer},
  {Reimer}, {Reposeur}, {Rodriguez}, {Roth}, {Sadrozinski}, {Sander}, {Saz
  Parkinson}, {Scargle}, {Sellerholm}, {Sgr{\`o}}, {Siskind}, {Smith},
  {Spandre}, {Spinelli}, {Starck}, {Strickman}, {Suson}, {Takahashi}, {Tanaka},
  {Thayer}, {Thayer}, {Torres}, {Uchiyama}, {Usher}, {Vasileiou}, {Vilchez},
  {Vitale}, {Waite}, {Wang}, {Winer}, {Wood}, {Ylinen}, {Zaharijas}, \&
  {Ziegler}}]{cosmowimp1}
{Abdo}, A.~A., {et~al.} 2010{\natexlab{a}}, JCAP, 4, 14

\bibitem[{{Abdo} {et~al.}(2010{\natexlab{b}}){Abdo}, {Ackermann}, {Ajello},
  {Allafort}, {Antolini}, {Atwood}, {Axelsson}, {Baldini}, {Ballet},
  {Barbiellini}, \& et~al.}]{1fgl}
---. 2010{\natexlab{b}}, \apjs, 188, 405

\bibitem[{{Abdo} {et~al.}(2010{\natexlab{c}}){Abdo}, {Ackermann}, {Ajello},
  {Antolini}, {Baldini}, {Ballet}, {Barbiellini}, {Bastieri}, {Baughman},
  {Bechtol}, {Bellazzini}, {Berenji}, {Blandford}, {Bloom}, {Bonamente},
  {Borgland}, {Bouvier}, {Bregeon}, {Brez}, {Brigida}, {Bruel}, {Burnett},
  {Buson}, {Caliandro}, {Cameron}, {Caraveo}, {Carrigan}, {Casandjian},
  {Cavazzuti}, {Cecchi}, {{\c C}elik}, {Charles}, {Chekhtman}, {Cheung},
  {Chiang}, {Ciprini}, {Claus}, {Cohen-Tanugi}, {Conrad}, {Costamante},
  {Cutini}, {Dermer}, {de Angelis}, {de Palma}, {Silva}, {Drell}, {Dubois},
  {Dumora}, {Farnier}, {Favuzzi}, {Fegan}, {Focke}, {Fukazawa}, {Funk},
  {Fusco}, {Gargano}, {Gasparrini}, {Gehrels}, {Germani}, {Giglietto},
  {Giommi}, {Giordano}, {Glanzman}, {Godfrey}, {Grenier}, {Grove}, {Guiriec},
  {Hadasch}, {Hayashida}, {Hays}, {Healey}, {Horan}, {Hughes}, {Itoh},
  {J{\'o}hannesson}, {Johnson}, {Johnson}, {Johnson}, {Kamae}, {Katagiri},
  {Kataoka}, {Kawai}, {Kn{\"o}dlseder}, {Kuss}, {Lande}, {Latronico}, {Lee},
  {Lemoine-Goumard}, {Llena Garde}, {Longo}, {Loparco}, {Lott}, {Lovellette},
  {Lubrano}, {Madejski}, {Makeev}, {Mazziotta}, {McConville}, {McEnery},
  {Meurer}, {Michelson}, {Mitthumsiri}, {Mizuno}, {Monte}, {Monzani},
  {Morselli}, {Moskalenko}, {Murgia}, {Nolan}, {Norris}, {Nuss}, {Ohsugi},
  {Omodei}, {Orlando}, {Ormes}, {Ozaki}, {Paneque}, {Panetta}, {Parent},
  {Pelassa}, {Pepe}, {Pesce-Rollins}, {Piron}, {Porter}, {Rain{\`o}}, {Rando},
  {Razzano}, {Reimer}, {Reimer}, {Ritz}, {Rochester}, {Rodriguez}, {Romani},
  {Roth}, {Sadrozinski}, {Sander}, {Saz Parkinson}, {Scargle}, {Sgr{\`o}},
  {Shaw}, {Smith}, {Spandre}, {Spinelli}, {Starck}, {Strickman}, {Strong},
  {Suson}, {Tajima}, {Takahashi}, {Takahashi}, {Tanaka}, {Thayer}, {Thayer},
  {Thompson}, {Tibaldo}, {Torres}, {Tosti}, {Tramacere}, {Uchiyama}, {Usher},
  {Vasileiou}, {Vilchez}, {Vitale}, {Waite}, {Wang}, {Winer}, {Wood}, {Yang},
  {Ylinen}, {Ziegler}, \& {Fermi-LAT Collaboration}}]{pop_pap}
---. 2010{\natexlab{c}}, ApJ, 720, 435

\bibitem[{{Abdo} {et~al.}(2010{\natexlab{d}}){Abdo}, {Ackermann}, {Ajello},
  {Allafort}, {Antolini}, {Atwood}, {Axelsson}, {Baldini}, {Ballet},
  {Barbiellini}, {Bastieri}, {Baughman}, {Bechtol}, {Bellazzini}, {Berenji},
  {Blandford}, {Bloom}, {Bogart}, {Bonamente}, {Borgland}, {Bouvier},
  {Bregeon}, {Brez}, {Brigida}, {Bruel}, {Buehler}, {Burnett}, {Buson},
  {Caliandro}, {Cameron}, {Cannon}, {Caraveo}, {Carrigan}, {Casandjian},
  {Cavazzuti}, {Cecchi}, {{\c C}elik}, {Celotti}, {Charles}, {Chekhtman},
  {Chen}, {Cheung}, {Chiang}, {Ciprini}, {Claus}, {Cohen-Tanugi}, {Conrad},
  {Costamante}, {Cotter}, {Cutini}, {D'Elia}, {Dermer}, {de Angelis}, {de
  Palma}, {De Rosa}, {Digel}, {Silva}, {Drell}, {Dubois}, {Dumora}, {Escande},
  {Farnier}, {Favuzzi}, {Fegan}, {Ferrara}, {Focke}, {Fortin}, {Frailis},
  {Fukazawa}, {Funk}, {Fusco}, {Gargano}, {Gasparrini}, {Gehrels}, {Germani},
  {Giebels}, {Giglietto}, {Giommi}, {Giordano}, {Giroletti}, {Glanzman},
  {Godfrey}, {Grandi}, {Grenier}, {Grondin}, {Grove}, {Guiriec}, {Hadasch},
  {Harding}, {Hayashida}, {Hays}, {Healey}, {Hill}, {Horan}, {Hughes},
  {Iafrate}, {Itoh}, {J{\'o}hannesson}, {Johnson}, {Johnson}, {Johnson},
  {Johnson}, {Kamae}, {Katagiri}, {Kataoka}, {Kawai}, {Kerr}, {Kn{\"o}dlseder},
  {Kuss}, {Lande}, {Latronico}, {Lavalley}, {Lemoine-Goumard}, {Llena Garde},
  {Longo}, {Loparco}, {Lott}, {Lovellette}, {Lubrano}, {Madejski}, {Makeev},
  {Malaguti}, {Massaro}, {Mazziotta}, {McConville}, {McEnery}, {McGlynn},
  {Michelson}, {Mitthumsiri}, {Mizuno}, {Moiseev}, {Monte}, {Monzani},
  {Morselli}, {Moskalenko}, {Murgia}, {Nolan}, {Norris}, {Nuss}, {Ohno},
  {Ohsugi}, {Omodei}, {Orlando}, {Ormes}, {Ozaki}, {Paneque}, {Panetta},
  {Parent}, {Pelassa}, {Pepe}, {Pesce-Rollins}, {Piranomonte}, {Piron},
  {Porter}, {Rain{\`o}}, {Rando}, {Razzano}, {Reimer}, {Reimer}, {Reposeur},
  {Ripken}, {Ritz}, {Rodriguez}, {Romani}, {Roth}, {Ryde}, {Sadrozinski},
  {Sanchez}, {Sander}, {Saz Parkinson}, {Scargle}, {Sgr{\`o}}, {Shaw},
  {Siskind}, {Smith}, {Spandre}, {Spinelli}, {Starck}, {Stawarz}, {Strickman},
  {Suson}, {Tajima}, {Takahashi}, {Takahashi}, {Tanaka}, {Taylor}, {Thayer},
  {Thayer}, {Thompson}, {Tibaldo}, {Torres}, {Tosti}, {Tramacere}, {Ubertini},
  {Uchiyama}, {Usher}, {Vasileiou}, {Vilchez}, {Villata}, {Vitale}, {Waite},
  {Wallace}, {Wang}, {Winer}, {Wood}, {Yang}, {Ylinen}, \& {Ziegler}}]{agn_cat}
---. 2010{\natexlab{d}}, \apj, 715, 429

\bibitem[{{Abdo} {et~al.}(2011{\natexlab{a}}){Abdo}, {Ackermann}, {Ajello},
  {Baldini}, {Ballet}, {Barbiellini}, {Bastieri}, {Bechtol}, {Bellazzini},
  {Berenji}, \& et~al.}]{lat_421}
---. 2011{\natexlab{a}}, \apj, 736, 131

\bibitem[{{Abdo} {et~al.}(2011{\natexlab{b}}){Abdo}, {Ackermann}, {Ajello},
  {Allafort}, {Baldini}, {Ballet}, {Barbiellini}, {Baring}, {Bastieri},
  {Bechtol}, \& et~al.}]{lat_501}
---. 2011{\natexlab{b}}, \apj, 727, 129

\bibitem[{{Abramowski} {et~al.}(2011){Abramowski}, {Acero}, {Aharonian},
  {Akhperjanian}, {Anton}, {Barnacka}, {Barres de Almeida}, {Bazer-Bachi},
  {Becherini}, {Becker}, {Behera}, {Bernl{\"o}hr}, {Bochow}, {Boisson},
  {Bolmont}, {Bordas}, {Borrel}, {Brucker}, {Brun}, {Brun}, {Bulik},
  {B{\"u}sching}, {Carrigan}, {Casanova}, {Cerruti}, {Chadwick}, {Charbonnier},
  {Chaves}, {Cheesebrough}, {Chounet}, {Clapson}, {Coignet}, {Conrad},
  {Dalton}, {Daniel}, {Davids}, {Degrange}, {Deil}, {Dickinson},
  {Djannati-Ata{\"i}}, {Domainko}, {Drury}, {Dubois}, {Dubus}, {Dyks}, {Dyrda},
  {Egberts}, {Eger}, {Espigat}, {Fallon}, {Farnier}, {Fegan}, {Feinstein},
  {Fernandes}, {Fiasson}, {Fontaine}, {F{\"o}rster}, {F{\"u}{\ss}ling},
  {Gallant}, {Gast}, {G{\'e}rard}, {Gerbig}, {Giebels}, {Glicenstein},
  {Gl{\"u}ck}, {Goret}, {G{\"o}ring}, {Hague}, {Hampf}, {Hauser}, {Heinz},
  {Heinzelmann}, {Henri}, {Hermann}, {Hinton}, {Hoffmann}, {Hofmann},
  {Hofverberg}, {Horns}, {Jacholkowska}, {de Jager}, {Jahn}, {Jamrozy}, {Jung},
  {Kastendieck}, {Katarzy{\'n}ski}, {Katz}, {Kaufmann}, {Keogh}, {Kerschhaggl},
  {Khangulyan}, {Kh{\'e}lifi}, {Klochkov}, {Klu{\'z}niak}, {Kneiske}, {Komin},
  {Kosack}, {Kossakowski}, {Laffon}, {Lamanna}, {Lennarz}, {Lohse}, {Lopatin},
  {Lu}, {Marandon}, {Marcowith}, {Masbou}, {Maurin}, {Maxted}, {McComb},
  {Medina}, {M{\'e}hault}, {Moderski}, {Moulin}, {Naumann}, {Naumann-Godo}, {de
  Naurois}, {Nedbal}, {Nekrassov}, {Nguyen}, {Nicholas}, {Niemiec}, {Nolan},
  {Ohm}, {Olive}, {de O{\~n}a Wilhelmi}, {Opitz}, {Ostrowski}, {Panter}, {Paz
  Arribas}, {Pedaletti}, {Pelletier}, {Petrucci}, {Pita}, {P{\"u}hlhofer},
  {Punch}, {Quirrenbach}, {Raue}, {Rayner}, {Reimer}, {Reimer}, {Renaud}, {de
  Los Reyes}, {Rieger}, {Ripken}, {Rob}, {Rosier-Lees}, {Rowell}, {Rudak},
  {Rulten}, {Ruppel}, {Ryde}, {Sahakian}, {Santangelo}, {Schlickeiser},
  {Sch{\"o}ck}, {Sch{\"o}nwald}, {Schwanke}, {Schwarzburg}, {Schwemmer},
  {Shalchi}, {Sikora}, {Skilton}, {Sol}, {Spengler}, {Stawarz}, {Steenkamp},
  {Stegmann}, {Stinzing}, {Sushch}, {Szostek}, {Tavernet}, {Terrier},
  {Tibolla}, {Tluczykont}, {Valerius}, {van Eldik}, {Vasileiadis}, {Venter},
  {Vialle}, {Viana}, {Vincent}, {Vivier}, {V{\"o}lk}, {Volpe}, {Vorobiov},
  {Vorster}, {Wagner}, {Ward}, {Wierzcholska}, {Zajczyk}, {Zdziarski}, {Zech},
  \& {Zechlin}}]{hess_gc11}
{Abramowski}, A., {et~al.} 2011, Physical Review Letters, 106, 161301

\bibitem[{Ackermann {et~al.}(2012)Ackermann, Ajello, Albert, Baldini, Ballet,
  Barbiellini, Bastieri, Bechtol, Bellazzini, Bloom, Bonamente, Borgland,
  Brandt, Bregeon, Brigida, Bruel, Buehler, Buson, Caliandro, Cameron, Caraveo,
  Cecchi, Charles, Chekhtman, Chiang, Ciprini, Claus, Cohen-Tanugi, Conrad,
  Cuoco, Cutini, D'ammando, de~Palma, Dermer, Digel, do~Couto~e Silva, Drell,
  Drlica-Wagner, Dubois, Favuzzi, Fegan, Ferrara, Fortin, Fukazawa, Fusco,
  Gargano, Gasparrini, Germani, Giglietto, Giroletti, Glanzman, Godfrey,
  Gomez-Vargas, Gr{\'e}goire, Grenier, Grove, Guiriec, Gustafsson, Hadasch,
  Hayashida, Hayashi, Hou, Hughes, J{\'o}hannesson, Johnson, Kamae,
  Kn{\"o}dlseder, Kuss, Lande, Latronico, Lemoine-Goumard, Linden, Lionetto,
  Llena~Garde, Longo, Loparco, Lovellette, Lubrano, Mazziotta, McEnery,
  Mitthumsiri, Mizuno, Monte, Monzani, Morselli, Moskalenko, Murgia,
  Naumann-Godo, Norris, Nuss, Ohsugi, Okumura, Orienti, Orlando, Ormes,
  Paneque, Panetta, Parent, Pavlidou, Pesce-Rollins, Pierbattista, Piron,
  Pivato, Rain{\`o}, Rando, Reimer, Reimer, Roth, Sbarra, Schmitt, Sgr{\`o},
  Siegal-Gaskins, Siskind, Spandre, Spinelli, Strong, Suson, Takahashi, Tanaka,
  Thayer, Tibaldo, Tinivella, Torres, Tosti, Troja, Usher, Vandenbroucke,
  Vasileiou, Vianello, Vitale, Waite, Winer, Wood, Wood, Yang, Zimmer, Komatsu,
  \& {Fermi LAT Collaboration}}]{aniso2012}
Ackermann, M., {et~al.} 2012, Physical Review D, 85, 083007

\bibitem[{{Ackermann} {et~al.}(2012){Ackermann}, {Ajello}, {Allafort},
  {Baldini}, {Ballet}, {Bastieri}, {Bechtol}, {Bellazzini}, {Berenji}, {Bloom},
  {Bonamente}, {Borgland}, {Bouvier}, {Bregeon}, {Brigida}, {Bruel}, {Buehler},
  {Buson}, {Caliandro}, {Cameron}, {Caraveo}, {Casandjian}, {Cecchi},
  {Charles}, {Chekhtman}, {Cheung}, {Chiang}, {Cillis}, {Ciprini}, {Claus},
  {Cohen-Tanugi}, {Conrad}, {Cutini}, {de Palma}, {Dermer}, {Digel}, {Silva},
  {Drell}, {Drlica-Wagner}, {Favuzzi}, {Fegan}, {Fortin}, {Fukazawa}, {Funk},
  {Fusco}, {Gargano}, {Gasparrini}, {Germani}, {Giglietto}, {Giordano},
  {Glanzman}, {Godfrey}, {Grenier}, {Guiriec}, {Gustafsson}, {Hadasch},
  {Hayashida}, {Hays}, {Hughes}, {J{\'o}hannesson}, {Johnson}, {Kamae},
  {Katagiri}, {Kataoka}, {Kn{\"o}dlseder}, {Kuss}, {Lande}, {Longo}, {Loparco},
  {Lott}, {Lovellette}, {Lubrano}, {Madejski}, {Martin}, {Mazziotta},
  {McEnery}, {Michelson}, {Mizuno}, {Monte}, {Monzani}, {Morselli},
  {Moskalenko}, {Murgia}, {Nishino}, {Norris}, {Nuss}, {Ohno}, {Ohsugi},
  {Okumura}, {Omodei}, {Orlando}, {Ozaki}, {Parent}, {Persic}, {Pesce-Rollins},
  {Petrosian}, {Pierbattista}, {Piron}, {Pivato}, {Porter}, {Rain{\`o}},
  {Rando}, {Razzano}, {Reimer}, {Reimer}, {Ritz}, {Roth}, {Sbarra}, {Sgr{\`o}},
  {Siskind}, {Spandre}, {Spinelli}, {Stawarz}, {Strong}, {Takahashi}, {Tanaka},
  {Thayer}, {Tibaldo}, {Tinivella}, {Torres}, {Tosti}, {Troja}, {Uchiyama},
  {Vandenbroucke}, {Vianello}, {Vitale}, {Waite}, {Wood}, \&
  {Yang}}]{lat_starforming}
{Ackermann}, M., {et~al.} 2012, \apj, 755, 164

\bibitem[{{Ackermann} {et~al.}(2013){Ackermann}, {Ajello}, {Allafort},
  {Atwood}, {Baldini}, {Ballet}, {Barbiellini}, {Bastieri}, {Bechtol},
  {Belfiore}, {Bellazzini}, {Bernieri}, {Bissaldi}, {Bloom}, {Bonamente},
  {Brandt}, {Bregeon}, {Brigida}, {Bruel}, {Buehler}, {Burnett}, {Buson},
  {Caliandro}, {Cameron}, {Campana}, {Caraveo}, {Casandjian}, {Cavazzuti},
  {Cecchi}, {Charles}, {Chaves}, {Chekhtman}, {Cheung}, {Chiang}, {Chiaro},
  {Ciprini}, {Claus}, {Cohen-Tanugi}, {Cominsky}, {Conrad}, {Cutini},
  {D'Ammando}, {de Angelis}, {de Palma}, {Dermer}, {Desiante}, {Digel}, {Di
  Venere}, {Drell}, {Drlica-Wagner}, {Favuzzi}, {Fegan}, {Ferrara}, {Focke},
  {Fortin}, {Franckowiak}, {Funk}, {Fusco}, {Gargano}, {Gasparrini}, {Gehrels},
  {Germani}, {Giglietto}, {Giommi}, {Giordano}, {Giroletti}, {Godfrey},
  {Gomez-Vargas}, {Grenier}, {Guiriec}, {Hadasch}, {Hanabata}, {Harding},
  {Hayashida}, {Hays}, {Hewitt}, {Hill}, {Horan}, {Hughes}, {Jogler},
  {J{\'o}hannesson}, {Johnson}, {Johnson}, {Johnson}, {Kamae}, {Kataoka},
  {Kawano}, {Kn{\"o}dlseder}, {Kuss}, {Lande}, {Larsson}, {Latronico},
  {Lemoine-Goumard}, {Longo}, {Loparco}, {Lott}, {Lovellette}, {Lubrano},
  {Massaro}, {Mayer}, {Mazziotta}, {McEnery}, {Mehault}, {Michelson}, {Mizuno},
  {Moiseev}, {Monzani}, {Morselli}, {Moskalenko}, {Murgia}, {Nemmen}, {Nuss},
  {Ohsugi}, {Okumura}, {Orienti}, {Ormes}, {Paneque}, {Perkins},
  {Pesce-Rollins}, {Piron}, {Pivato}, {Porter}, {Rain{\`o}}, {Razzano},
  {Reimer}, {Reimer}, {Reposeur}, {Ritz}, {Romani}, {Roth}, {Saz Parkinson},
  {Schulz}, {Sgr{\`o}}, {Siskind}, {Smith}, {Spandre}, {Spinelli}, {Stawarz},
  {Strong}, {Suson}, {Takahashi}, {Thayer}, {Thayer}, {Thompson}, {Tibaldo},
  {Tinivella}, {Torres}, {Tosti}, {Troja}, {Uchiyama}, {Usher},
  {Vandenbroucke}, {Vasileiou}, {Vianello}, {Vitale}, {Werner}, {Winer},
  {Wood}, \& {Wood}}]{1FHL}
---. 2013, \apjs, 209, 34

\bibitem[{{Ackermann} {et~al.}(2014{\natexlab{a}}){Ackermann}, {Albert},
  {Anderson}, {Baldini}, {Ballet}, {Barbiellini}, {Bastieri}, {Bechtol},
  {Bellazzini}, {Bissaldi}, {Bloom}, {Bonamente}, {Bouvier}, {Brandt},
  {Bregeon}, {Brigida}, {Bruel}, {Buehler}, {Buson}, {Caliandro}, {Cameron},
  {Caragiulo}, {Caraveo}, {Cecchi}, {Charles}, {Chekhtman}, {Chiang},
  {Ciprini}, {Claus}, {Cohen-Tanugi}, {Conrad}, {D'Ammando}, {de Angelis},
  {Dermer}, {Digel}, {do Couto e Silva}, {Drell}, {Drlica-Wagner}, {Essig},
  {Favuzzi}, {Ferrara}, {Franckowiak}, {Fukazawa}, {Funk}, {Fusco}, {Gargano},
  {Gasparrini}, {Giglietto}, {Giroletti}, {Godfrey}, {Gomez-Vargas}, {Grenier},
  {Guiriec}, {Gustafsson}, {Hayashida}, {Hays}, {Hewitt}, {Hughes}, {Jogler},
  {Kamae}, {Kn{\"o}dlseder}, {Kocevski}, {Kuss}, {Larsson}, {Latronico}, {Llena
  Garde}, {Longo}, {Loparco}, {Lovellette}, {Lubrano}, {Martinez}, {Mayer},
  {Mazziotta}, {Michelson}, {Mitthumsiri}, {Mizuno}, {Moiseev}, {Monzani},
  {Morselli}, {Moskalenko}, {Murgia}, {Nemmen}, {Nuss}, {Ohsugi}, {Orlando},
  {Ormes}, {Perkins}, {Piron}, {Pivato}, {Porter}, {Rain{\`o}}, {Rando},
  {Razzano}, {Razzaque}, {Reimer}, {Reimer}, {Ritz}, {S{\'a}nchez-Conde},
  {Sehgal}, {Sgr{\`o}}, {Siskind}, {Spinelli}, {Strigari}, {Suson}, {Tajima},
  {Takahashi}, {Thayer}, {Tibaldo}, {Tinivella}, {Torres}, {Uchiyama}, {Usher},
  {Vandenbroucke}, {Vianello}, {Vitale}, {Werner}, {Winer}, {Wood}, {Wood},
  {Zaharijas}, {Zimmer}, \& {Fermi-LAT Collaboration}}]{lat_dwarfs14}
---. 2014{\natexlab{a}}, \prd, 89, 042001

\bibitem[{{Ackermann} {et~al.}(2014{\natexlab{b}}){Ackermann}, {Ajello},
  {Albert}, {Atwood}, {Baldini}, {Ballet}, {Barbiellini}, {Bastieri},
  {Bechtol}, {Bellazzini}, {Bissaldi}, {Blandford}, {Bloom}, {Bottacini},
  {Brandt}, {Bregeon}, {Bruel}, {Buehler}, {Buson}, {Caliandro}, {Cameron},
  {Caragiulo}, {Caraveo}, {Cavazzuti}, {Cecchi}, {Charles}, {Chekhtman},
  {Chiang}, {Chiaro}, {Ciprini}, {Claus}, {Cohen-Tanugi}, {Conrad}, {Cuoco},
  {Cutini}, {D'Ammando}, {de Angelis}, {de Palma}, {Dermer}, {Digel}, {Silva},
  {Drell}, {Favuzzi}, {Ferrara}, {Focke}, {Franckowiak}, {Fukazawa}, {Funk},
  {Fusco}, {Gargano}, {Gasparrini}, {Germani}, {Giglietto}, {Giommi},
  {Giordano}, {Giroletti}, {Godfrey}, {Gomez-Vargas}, {Grenier}, {Guiriec},
  {Gustafsson}, {Hadasch}, {Hayashi}, {Hays}, {Hewitt}, {Ippoliti}, {Jogler},
  {J{\'o}hannesson}, {Johnson}, {Johnson}, {Kamae}, {Kataoka},
  {Kn{\"o}dlseder}, {Kuss}, {Larsson}, {Latronico}, {Li}, {Li}, {Longo},
  {Loparco}, {Lott}, {Lovellette}, {Lubrano}, {Madejski}, {Manfreda},
  {Massaro}, {Mayer}, {Mazziotta}, {McEnery}, {Michelson}, {Mitthumsiri},
  {Mizuno}, {Moiseev}, {Monzani}, {Morselli}, {Moskalenko}, {Murgia}, {Nemmen},
  {Nuss}, {Ohsugi}, {Omodei}, {Orlando}, {Ormes}, {Paneque}, {Panetta},
  {Perkins}, {Pesce-Rollins}, {Piron}, {Pivato}, {Porter}, {Rain{\`o}},
  {Rando}, {Razzano}, {Razzaque}, {Reimer}, {Reimer}, {Reposeur}, {Ritz},
  {Romani}, {S{\'a}nchez-Conde}, {Schaal}, {Schulz}, {Sgr{\`o}}, {Siskind},
  {Spandre}, {Spinelli}, {Strong}, {Suson}, {Takahashi}, {Thayer}, {Thayer},
  {Tibaldo}, {Tinivella}, {Torres}, {Tosti}, {Troja}, {Uchiyama}, {Vianello},
  {Werner}, {Winer}, {Wood}, {Wood}, {Zaharijas}, \& {Zimmer}}]{lat_egb2}
---. 2014{\natexlab{b}}, arXiv:1410.3696

\bibitem[{{Ackermann} {et~al.}(2014{\natexlab{c}}){Ackermann}, {Ajello},
  {Albert}, {Atwood}, {Baldini}, {Ballet}, {Barbiellini}, {Bastieri},
  {Bechtol}, {Bellazzini}, {Bissaldi}, {Blandford}, {Bloom}, {Bottacini},
  {Brandt}, {Bregeon}, {Bruel}, {Buehler}, {Buson}, {Caliandro}, {Cameron},
  {Caragiulo}, {Caraveo}, {Cavazzuti}, {Cecchi}, {Charles}, {Chekhtman},
  {Chiang}, {Chiaro}, {Ciprini}, {Claus}, {Cohen-Tanugi}, {Conrad}, {Cuoco},
  {Cutini}, {D'Ammando}, {de Angelis}, {de Palma}, {Dermer}, {Digel}, {Silva},
  {Drell}, {Favuzzi}, {Ferrara}, {Focke}, {Franckowiak}, {Fukazawa}, {Funk},
  {Fusco}, {Gargano}, {Gasparrini}, {Germani}, {Giglietto}, {Giommi},
  {Giordano}, {Giroletti}, {Godfrey}, {Gomez-Vargas}, {Grenier}, {Guiriec},
  {Gustafsson}, {Hadasch}, {Hayashi}, {Hays}, {Hewitt}, {Ippoliti}, {Jogler},
  {J{\'o}hannesson}, {Johnson}, {Johnson}, {Kamae}, {Kataoka},
  {Kn{\"o}dlseder}, {Kuss}, {Larsson}, {Latronico}, {Li}, {Li}, {Longo},
  {Loparco}, {Lott}, {Lovellette}, {Lubrano}, {Madejski}, {Manfreda},
  {Massaro}, {Mayer}, {Mazziotta}, {McEnery}, {Michelson}, {Mitthumsiri},
  {Mizuno}, {Moiseev}, {Monzani}, {Morselli}, {Moskalenko}, {Murgia}, {Nemmen},
  {Nuss}, {Ohsugi}, {Omodei}, {Orlando}, {Ormes}, {Paneque}, {Panetta},
  {Perkins}, {Pesce-Rollins}, {Piron}, {Pivato}, {Porter}, {Rain{\`o}},
  {Rando}, {Razzano}, {Razzaque}, {Reimer}, {Reimer}, {Reposeur}, {Ritz},
  {Romani}, {S{\'a}nchez-Conde}, {Schaal}, {Schulz}, {Sgr{\`o}}, {Siskind},
  {Spandre}, {Spinelli}, {Strong}, {Suson}, {Takahashi}, {Thayer}, {Thayer},
  {Tibaldo}, {Tinivella}, {Torres}, {Tosti}, {Troja}, {Uchiyama}, {Vianello},
  {Werner}, {Winer}, {Wood}, {Wood}, {Zaharijas}, \& {Zimmer}}]{cosmowimp2}
---. 2014{\natexlab{c}}, in prep.



\bibitem[{{Ackermann} {et~al.}(2014{\natexlab{d}}){Ackermann}, {Ajello},
  {Albert}, {Atwood}, {Baldini}, {Ballet}, {Barbiellini}, {Bastieri},
  {Bechtol}, {Bellazzini}, {Bissaldi}, {Blandford}, {Bloom}, {Bottacini},
  {Brandt}, {Bregeon}, {Bruel}, {Buehler}, {Buson}, {Caliandro}, {Cameron},
  {Caragiulo}, {Caraveo}, {Cavazzuti}, {Cecchi}, {Charles}, {Chekhtman},
  {Chiang}, {Chiaro}, {Ciprini}, {Claus}, {Cohen-Tanugi}, {Conrad}, {Cuoco},
  {Cutini}, {D'Ammando}, {de Angelis}, {de Palma}, {Dermer}, {Digel}, {Silva},
  {Drell}, {Favuzzi}, {Ferrara}, {Focke}, {Franckowiak}, {Fukazawa}, {Funk},
  {Fusco}, {Gargano}, {Gasparrini}, {Germani}, {Giglietto}, {Giommi},
  {Giordano}, {Giroletti}, {Godfrey}, {Gomez-Vargas}, {Grenier}, {Guiriec},
  {Gustafsson}, {Hadasch}, {Hayashi}, {Hays}, {Hewitt}, {Ippoliti}, {Jogler},
  {J{\'o}hannesson}, {Johnson}, {Johnson}, {Kamae}, {Kataoka},
  {Kn{\"o}dlseder}, {Kuss}, {Larsson}, {Latronico}, {Li}, {Li}, {Longo},
  {Loparco}, {Lott}, {Lovellette}, {Lubrano}, {Madejski}, {Manfreda},
  {Massaro}, {Mayer}, {Mazziotta}, {McEnery}, {Michelson}, {Mitthumsiri},
  {Mizuno}, {Moiseev}, {Monzani}, {Morselli}, {Moskalenko}, {Murgia}, {Nemmen},
  {Nuss}, {Ohsugi}, {Omodei}, {Orlando}, {Ormes}, {Paneque}, {Panetta},
  {Perkins}, {Pesce-Rollins}, {Piron}, {Pivato}, {Porter}, {Rain{\`o}},
  {Rando}, {Razzano}, {Razzaque}, {Reimer}, {Reimer}, {Reposeur}, {Ritz},
  {Romani}, {S{\'a}nchez-Conde}, {Schaal}, {Schulz}, {Sgr{\`o}}, {Siskind},
  {Spandre}, {Spinelli}, {Strong}, {Suson}, {Takahashi}, {Thayer}, {Thayer},
  {Tibaldo}, {Tinivella}, {Torres}, {Tosti}, {Troja}, {Uchiyama}, {Vianello},
  {Werner}, {Winer}, {Wood}, {Wood}, {Zaharijas}, \& {Zimmer}}]{3LAC}
---. 2014{\natexlab{d}}, in prep.

\bibitem[{{Ahlers} \& {Salvado}(2011)}]{ahlers11}
{Ahlers}, M., \& {Salvado}, J. 2011, \prd, 84, 085019

\bibitem[{{Ajello} {et~al.}(2014){Ajello}, {Romani}, {Gasparrini}, {Shaw},
  {Bolmer}, {Cotter}, {Finke}, {Greiner}, {Healey}, {King}, {Max-Moerbeck},
  {Michelson}, {Potter}, {Rau}, {Readhead}, {Richards}, \& {Schady}}]{ajello14}
{Ajello}, M., {et~al.} 2014, \apj, 780, 73

\bibitem[{{Aleksi{\'c}} {et~al.}(2014){Aleksi{\'c}}, {Ansoldi}, {Antonelli},
  {Antoranz}, {Babic}, {Bangale}, {Barres de Almeida}, {Barrio}, {Becerra
  Gonz{\'a}lez}, {Bednarek}, {Berger}, {Bernardini}, {Biland}, {Blanch},
  {Bock}, {Bonnefoy}, {Bonnoli}, {Borracci}, {Bretz}, {Carmona}, {Carosi},
  {Carreto Fidalgo}, {Colin}, {Colombo}, {Contreras}, {Cortina}, {Covino}, {Da
  Vela}, {Dazzi}, {De Angelis}, {De Caneva}, {De Lotto}, {Delgado Mendez},
  {Doert}, {Dom{\'{\i}}nguez}, {Dominis Prester}, {Dorner}, {Doro}, {Einecke},
  {Eisenacher}, {Elsaesser}, {Farina}, {Ferenc}, {Fonseca}, {Font}, {Frantzen},
  {Fruck}, {Garc{\'{\i}}a L{\'o}pez}, {Garczarczyk}, {Garrido Terrats}, {Gaug},
  {Giavitto}, {Godinovi{\'c}}, {Gonz{\'a}lez Mu{\~n}oz}, {Gozzini}, {Hadasch},
  {Hayashida}, {Herrero}, {Hildebrand}, {Hose}, {Hrupec}, {Idec}, {Kadenius},
  {Kellermann}, {Kodani}, {Konno}, {Krause}, {Kubo}, {Kushida}, {La Barbera},
  {Lelas}, {Lewandowska}, {Lindfors}, {Lombardi}, {L{\'o}pez},
  {L{\'o}pez-Coto}, {L{\'o}pez-Oramas}, {Lorenz}, {Lozano}, {Makariev},
  {Mallot}, {Maneva}, {Mankuzhiyil}, {Mannheim}, {Maraschi}, {Marcote},
  {Mariotti}, {Mart{\'{\i}}nez}, {Mazin}, {Menzel}, {Meucci}, {Miranda},
  {Mirzoyan}, {Moralejo}, {Munar-Adrover}, {Nakajima}, {Niedzwiecki},
  {Nilsson}, {Nishijima}, {Nowak}, {Orito}, {Overkemping}, {Paiano},
  {Palatiello}, {Paneque}, {Paoletti}, {Paredes}, {Paredes-Fortuny}, {Partini},
  {Persic}, {Prada}, {Prada Moroni}, {Prandini}, {Preziuso}, {Puljak},
  {Reinthal}, {Rhode}, {Rib{\'o}}, {Rico}, {Rodriguez Garcia}, {R{\"u}gamer},
  {Saggion}, {Saito}, {Saito}, {Salvati}, {Satalecka}, {Scalzotto}, {Scapin},
  {Schultz}, {Schweizer}, {Sillanp{\"a}{\"a}}, {Sitarek}, {Snidaric},
  {Sobczynska}, {Spanier}, {Stamatescu}, {Stamerra}, {Steinbring}, {Storz},
  {Sun}, {Suri{\'c}}, {Takalo}, {Takami}, {Tavecchio}, {Temnikov},
  {Terzi{\'c}}, {Tescaro}, {Teshima}, {Thaele}, {Tibolla}, {Torres}, {Toyama},
  {Treves}, {Uellenbeck}, {Vogler}, {Wagner}, {Zandanel}, {Zanin}, \&
  {Ibarra}}]{magic_segue14}
{Aleksi{\'c}}, J., {et~al.} 2014, \jcap, 2, 8

\bibitem[{{Aliu} {et~al.}(2012){Aliu}, {Archambault}, {Arlen}, {Aune},
  {Beilicke}, {Benbow}, {B{\"o}ttcher}, {Bouvier}, {Bradbury}, {Buckley},
  {Bugaev}, {Byrum}, {Cannon}, {Cesarini}, {Ciupik}, {Collins-Hughes},
  {Connolly}, {Coppi}, {Cui}, {Decerprit}, {Dickherber}, {Dumm}, {Errando},
  {Falcone}, {Feng}, {Finley}, {Finnegan}, {Fortson}, {Furniss}, {Galante},
  {Gall}, {Godambe}, {Griffin}, {Grube}, {Gyuk}, {Hanna}, {Hawkins}, {Holder},
  {Huan}, {Hughes}, {Humensky}, {Kaaret}, {Karlsson}, {Kertzman}, {Khassen},
  {Kieda}, {Krawczynski}, {Krennrich}, {Lang}, {Lee}, {Madhavan}, {Maier},
  {Majumdar}, {McArthur}, {McCann}, {Moriarty}, {Mukherjee}, {Ong}, {Orr},
  {Otte}, {Palma}, {Park}, {Perkins}, {Pichel}, {Pohl}, {Prokoph}, {Quinn},
  {Ragan}, {Reyes}, {Reynolds}, {Roache}, {Rose}, {Ruppel}, {Saxon},
  {Schroedter}, {Sembroski}, {{\c S}ent{\"u}rk}, {Smith}, {Staszak},
  {Telezhinsky}, {Te{\v s}i{\'c}}, {Theiling}, {Thibadeau}, {Tsurusaki},
  {Varlotta}, {Vivier}, {Wakely}, {Ward}, {Weekes}, {Weinstein}, {Weisgarber},
  {Williams}, {Zitzer}, {Fortin}, \& {Horan}}]{rbs0413}
{Aliu}, E., {et~al.} 2012, \apj, 750, 94

\bibitem[{{Atwood} {et~al.}(2009){Atwood}, {Abdo}, {Ackermann}, {Althouse},
  {Anderson}, {Axelsson}, {Baldini}, {Ballet}, {Band}, {Barbiellini},
  {Bartelt}, {Bastieri}, {Baughman}, {Bechtol}, {B{\'e}d{\'e}r{\`e}de},
  {Bellardi}, {Bellazzini}, {Berenji}, {Bignami}, {Bisello}, {Bissaldi},
  {Blandford}, {Bloom}, {Bogart}, {Bonamente}, {Bonnell}, {Borgland},
  {Bouvier}, {Bregeon}, {Brez}, {Brigida}, {Bruel}, {Burnett}, {Busetto},
  {Caliandro}, {Cameron}, {Caraveo}, {Carius}, {Carlson}, {Casandjian},
  {Cavazzuti}, {Ceccanti}, {Cecchi}, {Charles}, {Chekhtman}, {Cheung},
  {Chiang}, {Chipaux}, {Cillis}, {Ciprini}, {Claus}, {Cohen-Tanugi},
  {Condamoor}, {Conrad}, {Corbet}, {Corucci}, {Costamante}, {Cutini}, {Davis},
  {Decotigny}, {DeKlotz}, {Dermer}, {de Angelis}, {Digel}, {do Couto e Silva},
  {Drell}, {Dubois}, {Dumora}, {Edmonds}, {Fabiani}, {Farnier}, {Favuzzi},
  {Flath}, {Fleury}, {Focke}, {Funk}, {Fusco}, {Gargano}, {Gasparrini},
  {Gehrels}, {Gentit}, {Germani}, {Giebels}, {Giglietto}, {Giommi}, {Giordano},
  {Glanzman}, {Godfrey}, {Grenier}, {Grondin}, {Grove}, {Guillemot}, {Guiriec},
  {Haller}, {Harding}, {Hart}, {Hays}, {Healey}, {Hirayama}, {Hjalmarsdotter},
  {Horn}, {Hughes}, {J{\'o}hannesson}, {Johansson}, {Johnson}, {Johnson},
  {Johnson}, {Johnson}, {Kamae}, {Katagiri}, {Kataoka}, {Kavelaars}, {Kawai},
  {Kelly}, {Kerr}, {Klamra}, {Kn{\"o}dlseder}, {Kocian}, {Komin}, {Kuehn},
  {Kuss}, {Landriu}, {Latronico}, {Lee}, {Lee}, {Lemoine-Goumard}, {Lionetto},
  {Longo}, {Loparco}, {Lott}, {Lovellette}, {Lubrano}, {Madejski}, {Makeev},
  {Marangelli}, {Massai}, {Mazziotta}, {McEnery}, {Menon}, {Meurer},
  {Michelson}, {Minuti}, {Mirizzi}, {Mitthumsiri}, {Mizuno}, {Moiseev},
  {Monte}, {Monzani}, {Moretti}, {Morselli}, {Moskalenko}, {Murgia},
  {Nakamori}, {Nishino}, {Nolan}, {Norris}, {Nuss}, {Ohno}, {Ohsugi}, {Omodei},
  {Orlando}, {Ormes}, {Paccagnella}, {Paneque}, {Panetta}, {Parent}, {Pearce},
  {Pepe}, {Perazzo}, {Pesce-Rollins}, {Picozza}, {Pieri}, {Pinchera}, {Piron},
  {Porter}, {Poupard}, {Rain{\`o}}, {Rando}, {Rapposelli}, {Razzano}, {Reimer},
  {Reimer}, {Reposeur}, {Reyes}, {Ritz}, {Rochester}, {Rodriguez}, {Romani},
  {Roth}, {Russell}, {Ryde}, {Sabatini}, {Sadrozinski}, {Sanchez}, {Sander},
  {Sapozhnikov}, {Parkinson}, {Scargle}, {Schalk}, {Scolieri}, {Sgr{\`o}},
  {Share}, {Shaw}, {Shimokawabe}, {Shrader}, {Sierpowska-Bartosik}, {Siskind},
  {Smith}, {Smith}, {Spandre}, {Spinelli}, {Starck}, {Stephens}, {Strickman},
  {Strong}, {Suson}, {Tajima}, {Takahashi}, {Takahashi}, {Tanaka}, {Tenze},
  {Tether}, {Thayer}, {Thayer}, {Thompson}, {Tibaldo}, {Tibolla}, {Torres},
  {Tosti}, {Tramacere}, {Turri}, {Usher}, {Vilchez}, {Vitale}, {Wang},
  {Watters}, {Winer}, {Wood}, {Ylinen}, \& {Ziegler}}]{atwood09}
{Atwood}, W.~B., {et~al.} 2009, \apj, 697, 1071

\bibitem[{{Bertone} {et~al.}(2005){Bertone}, {Hooper}, \& {Silk}}]{bertone05}
{Bertone}, G., {Hooper}, D., \& {Silk}, J. 2005, \physrep, 405, 279

\bibitem[{Bhattacharjee \& Sigl(2000)}]{bhattacharjee00}
Bhattacharjee, P., \& Sigl, G. 2000, Physics Reports, 327, 109

\bibitem[{Bringmann {et~al.}(2014)Bringmann, Calore, Di~Mauro, \&
  Donato}]{calore2013}
Bringmann, T., Calore, F., Di~Mauro, M., \& Donato, F. 2014, Phys.Rev., D89,
  023012

\bibitem[{{Calore} {et~al.}(2014){Calore}, {Di Mauro}, \& {Donato}}]{calore14}
{Calore}, F., {Di Mauro}, M., \& {Donato}, F. 2014, arXiv:1406.2706

\bibitem[{{Cholis} {et~al.}(2014){Cholis}, {Hooper}, \& {McDermott}}]{cholis14}
{Cholis}, I., {Hooper}, D., \& {McDermott}, S.~D. 2014, \jcap, 2, 14

\bibitem[{Cuoco {et~al.}(2012)Cuoco, Komatsu, \& Siegal-Gaskins}]{cuoco12}
Cuoco, A., Komatsu, E., \& Siegal-Gaskins, J.~M. 2012, Physical Review D, 86,
  063004

\bibitem[{{Dermer}(2007)}]{dermer07}
{Dermer}, C.~D. 2007, \apj, 659, 958

\bibitem[{Di~Mauro {et~al.}(2013)Di~Mauro, Calore, Donato, Ajello, \&
  Latronico}]{dimauro13}
Di~Mauro, M., Calore, F., Donato, F., Ajello, M., \& Latronico, L. 2013, \apj,
  780, 161


\bibitem[{{Di Mauro} {et~al.}(2014){Di Mauro}, {Donato}, {Lamanna}, {Sanchez},
  \& {Serpico}}]{dimauro14}
{Di Mauro}, M., {Donato}, F., {Lamanna}, G., {Sanchez}, D.~A., \& {Serpico},
  P.~D. 2014, \apj, 786, 129


\bibitem[{{Fields} {et~al.}(2010){Fields}, {Pavlidou}, \&
  {Prodanovi{\'c}}}]{fields10}
{Fields}, B.~D., {Pavlidou}, V., \& {Prodanovi{\'c}}, T. 2010, \apjl, 722, L199

\bibitem[{Finke {et~al.}(2010)Finke, Razzaque, \& Dermer}]{finke10}
Finke, J.~D., Razzaque, S., \& Dermer, C.~D. 2010, \apj, 712, 238

\bibitem[{{Fornasa} {et~al.}(2013){Fornasa}, {Zavala}, {S{\'a}nchez-Conde},
  {Siegal-Gaskins}, {Delahaye}, {Prada}, {Vogelsberger}, {Zandanel}, \&
  {Frenk}}]{fornasa13}
{Fornasa}, M., {et~al.} 2013, \mnras, 429, 1529

\bibitem[{G\'{o}mez-Vargas {et~al.}(2014)G\'{o}mez-Vargas, Cuoco, Linden,
  Sánchez-Conde, Siegal-Gaskins, Delahaye, Fornasa, Komatsu, Prada, \&
  Zavala}]{german14}
G\'{o}mez-Vargas, G., {et~al.} 2014, Nuclear Instruments and Methods in Physics
  Research Section A: Accelerators, Spectrometers, Detectors and Associated
  Equipment, 742, 149 , 4th Roma International Conference on Astroparticle
  Physics

\bibitem[{{Gould} \& {Schr{\'e}der}(1966)}]{gould66}
{Gould}, R.~J., \& {Schr{\'e}der}, G. 1966, PRL, 16, 252

\bibitem[{{Harding} \& {Abazajian}(2012)}]{harding12}
{Harding}, J.~P., \& {Abazajian}, K.~N. 2012, \jcap, 11, 26

\bibitem[{{Inoue}(2011)}]{inoue11b}
{Inoue}, Y. 2011, \apj, 733, 66

\bibitem[{{Lacki} {et~al.}(2014){Lacki}, {Horiuchi}, \& {Beacom}}]{lacki14}
{Lacki}, B.~C., {Horiuchi}, S., \& {Beacom}, J.~F. 2014, \apj, 786, 40

\bibitem[{{Loeb} \& {Waxman}(2000)}]{loeb00}
{Loeb}, A., \& {Waxman}, E. 2000, \nat, 405, 156

\bibitem[{{Makiya} {et~al.}(2010){Makiya}, {Totani}, \&
  {Kobayashi}}]{makiya2010}
{Makiya}, R., {Totani}, T., \& {Kobayashi}, M.~A.~R. 2010, ArXiv:1005.1390

\bibitem[{{Miniati}(2002)}]{miniati02}
{Miniati}, F. 2002, \mnras, 337, 199

\bibitem[{{Nolan} {et~al.}(2012){Nolan}, {Abdo}, {Ackermann}, {Ajello},
  {Allafort}, {Antolini}, {Atwood}, {Axelsson}, {Baldini}, {Ballet}, \&
  et~al.}]{2fgl}
{Nolan}, P.~L., {et~al.} 2012, \apjs, 199, 31

\bibitem[{{S{\'a}nchez-Conde} \& {Prada}(2014)}]{masc14}
{S{\'a}nchez-Conde}, M.~A., \& {Prada}, F. 2014, \mnras, 442, 2271

\bibitem[{{Sefusatti} {et~al.}(2014){Sefusatti}, {Zaharijas}, {Serpico},
  {Theurel}, \& {Gustafsson}}]{sefusatti14}
{Sefusatti}, E., {Zaharijas}, G., {Serpico}, P.~D., {Theurel}, D., \&
  {Gustafsson}, M. 2014, \mnras, 441, 1861

\bibitem[{Serpico {et~al.}(2012)Serpico, Sefusatti, Gustafsson, \&
  Zaharijas}]{serpico2011}
Serpico, P.~D., Sefusatti, E., Gustafsson, M., \& Zaharijas, G. 2012,
  Mon.Not.Roy.Astron.Soc., 421, L87

\bibitem[{{Shaw} {et~al.}(2013){Shaw}, {Romani}, {Cotter}, {Healey},
  {Michelson}, {Readhead}, {Richards}, {Max-Moerbeck}, {King}, \&
  {Potter}}]{shaw13}
{Shaw}, M.~S., {et~al.} 2013, \apj, 764, 135

\bibitem[{{Singal} {et~al.}(2012){Singal}, {Petrosian}, \& {Ajello}}]{singal12}
{Singal}, J., {Petrosian}, V., \& {Ajello}, M. 2012, \apj, 753, 45

\bibitem[{{Stecker} {et~al.}(1992){Stecker}, {de Jager}, \&
  {Salamon}}]{stecker92}
{Stecker}, F.~W., {de Jager}, O.~C., \& {Salamon}, M.~H. 1992, \apjl, 390, L49

\bibitem[{{Stecker} \& {Venters}(2011)}]{stecker11}
{Stecker}, F.~W., \& {Venters}, T.~M. 2011, \apj, 736, 40

\bibitem[{{Thompson} {et~al.}(2007){Thompson}, {Quataert}, \&
  {Waxman}}]{thompson07}
{Thompson}, T.~A., {Quataert}, E., \& {Waxman}, E. 2007, \apj, 654, 219

\bibitem[{{Ullio} {et~al.}(2002){Ullio}, {Bergstr{\"o}m}, {Edsj{\"o}}, \&
  {Lacey}}]{Ullio02}
{Ullio}, P., {Bergstr{\"o}m}, L., {Edsj{\"o}}, J., \& {Lacey}, C. 2002, \prd,
  66, 123502

\end{thebibliography}

\end{document}